\documentclass{imsart}

\RequirePackage{amsthm,amsmath}
\RequirePackage{natbib}
\RequirePackage[colorlinks,citecolor=blue,urlcolor=blue]{hyperref}
\RequirePackage{graphicx}
\usepackage{amsmath,amsfonts}
\usepackage{array}
\usepackage{textcomp}
\usepackage{stfloats}
\usepackage{url}
\usepackage{verbatim}
\usepackage{graphicx}
\usepackage{bm}
\usepackage{algorithm}
\usepackage{algorithmic}
\usepackage{color}
\usepackage{xr}
\usepackage{float}
\usepackage{multirow}
\usepackage{amssymb}
\usepackage{amstext}

\newtheorem{assumption}{Assumption}
\newtheorem{lemma}{Lemma}
\newtheorem{theorem}{Theorem}

\newcommand*{\dif}{\mathop{}\!\mathrm{d}}

\startlocaldefs
\numberwithin{equation}{section}
\theoremstyle{plain}

\endlocaldefs

\begin{document}
	
	\begin{frontmatter}
		\title{A Spectral Approach for the Dynamic Bradley-Terry Model}

		\runtitle{Spectral Dynamic Ranking}
		
		\begin{aug}
			\author{\fnms{Xin-Yu} \snm{Tian}\thanksref{t1}\ead[label=e1]{tianxinyu@amss.ac.cn}}
			\and
			\author{\fnms{Jian} \snm{Shi}\thanksref{t1}\ead[label=e2]{jshi@iss.ac.cn}}
			
			\address{Academy of Mathematics and Systems Science,
				Chinese Academy of Sciences,\\
				Beijing 100190, 
				China\\ 
				School of Mathematical Sciences,
				University of Chinese Academy of Sciences,\\
				Beijing 100049, China\\
				\printead{e1,e2}}
			
			\author{\fnms{Xiaotong} \snm{Shen}\thanksref{t2}
				\ead[label=e3]{xshen@umn.edu}
			}
			
			\address{School of Statistics,
				The University of Minnesota,
				Minneapolis, MN 55455, USA\\
				\printead{e3}\\
			}
			
			\author{\fnms{Kai} \snm{Song}
				\ead[label=e4]{kaisong@amss.ac.cn}
			}
			
			\address{Academy of Mathematics and Systems Science,
				Chinese Academy of Sciences,\\
				Beijing 100190, 
				China\\
				\printead{e4}\\
			}
			
			\thankstext{t1}{Equal contribution.}
			\thankstext{t2}{Corresponding author.}
			\runauthor{X.Y. Tian et al.}
			
		\end{aug}
		
		\begin{abstract}
			The dynamic ranking, due to its increasing importance in many applications, is becoming crucial, especially with the collection of voluminous time-dependent data. One such application is sports statistics, where dynamic ranking aids in forecasting the performance of competitive teams, drawing on historical and current data. Despite its usefulness, predicting and inferring rankings pose challenges in environments necessitating time-dependent modeling. This paper introduces a spectral ranker called Kernel Rank Centrality, designed
			to rank items based on pairwise comparisons over time. The ranker operates via kernel smoothing in the Bradley-Terry model, utilizing a Markov chain model. Unlike the maximum likelihood approach, the spectral ranker is nonparametric, demands fewer model assumptions and computations, and allows for real-time ranking. We establish the asymptotic distribution of the ranker by applying an innovative group inverse technique, resulting
			in a uniform and precise entrywise expansion. This result allows us to devise a new inferential method for predictive inference, previously unavailable in existing approaches. Our numerical examples showcase the ranker’s utility in predictive accuracy and constructing an
			uncertainty measure for prediction, leveraging data from the National Basketball Association (NBA). The results underscore our method’s potential compared to the gold standard in sports, the Arpad Elo rating system.
			
		\end{abstract}
		
		
	\newpage
		\tableofcontents
	\end{frontmatter}

\section{Introduction}
\label{sec_intro}

The dynamic ranking has garnered interest across various fields such as sports \citep{Maystre2019}, engineering \citep{MAZZUCHI2008722}, recommendation systems \citep{ZHOU2019176}, and biology \citep{shah2017simple}. This technique involves ranking items over time based on established principles, like assigning a rating score to each item \citep{Baback2018, Liu2019model}. This paper emphasizes the prediction and inference of dynamic ranking.

The Bradley-Terry model \citep{BT1952}, a well-regarded model for static ranking, focuses on quantifying the winning probability of one team versus another in a sports tournament or the preference probability of a user towards an item in recommender systems across a user base. Its simplicity, practicality, and interpretability have contributed to its widespread adoption for analyzing rank data. Its dynamic counterpart, the time-varying Bradley-Terry model, enables the model probabilities to fluctuate over time, accommodating rankings at various time points. Despite this, the time-varying model has received less attention than its static counterpart, with the main challenge lying in predictive inference, especially uncertainty quantification of prediction accuracy \citep{liu2022lagrangian}. This paper aims to bridge this gap.

To treat the time-varying Bradley-Terry model, researchers have proposed two approaches: Bayesian and maximum likelihood. The Bayesian approach models the winning probability with a stochastic process like the Gaussian process \citep{Maystre2019, Glickman2001}, estimating model parameters by sampling from the posterior distribution through Markov chain Monte Carlo (MCMC). The maximum likelihood approach, as proposed by \cite{MCHALE2011619}, employs a time-weighted likelihood with an exponential decay weight function, whereas \cite{bong2020nonparametric} uses a likelihood method based on kernel smoothing. However, both approaches can be computationally intensive.

This paper introduces a dynamic ranker based on rank centrality and kernel, a spectral method utilizing pairwise comparisons, and a random walk on a directed graph. Our ranker dynamically ranks items using kernel smoothing, building upon the static ranking work of \cite{Negahban2017} on pairwise comparisons. This method offers the advantage of lower computational cost, facilitating analytic updating rather than optimization. \citet{karle2021dynamic} extends the method into the dynamic settings with a nearest neighbor strategy and discusses the convergence theory in a general graph case, which also motivates our method.
Theoretically, we establish the $\ell_{2}$ and $\ell_{\infty}$ convergence rates of the proposed ranker, providing an innovative method, which differs from the previous study \citep{karle2021dynamic}, for theoretical analysis. On this basis, we ascertain the asymptotic normality of the ranker estimator using a precise, uniform expansion allowing for a growing number of items. In contrast, the distribution theory for the maximum likelihood estimate of the time-varying Bradley-Terry model remains elusive, despite the results established by \cite{Simons1999} and \cite{Han2020} for the static Bradley-Terry model's maximum likelihood estimate.

The main contribution of this paper is the development and analysis of Kernel Rank Centrality (KRC), a spectral method for estimating item ratings in a dynamic setting using comparison data. KRC merges the Rank Centrality estimator with kernel smoothing to yield accurate and computationally efficient estimates. Theoretical results demonstrate that the KRC estimation converges to the dynamic Bradley-Terry model score with an $\ell_2$ error bound, with detailed results obtained for the asymptotic distribution and entrywise error. The proposed ranker's distribution theory is groundbreaking, especially the asymptotic normality determined using an innovative approximation of the group-inverse matrix. This result allows for a uniform expansion derivation for the rank estimator that can apply to static settings and sparse regimes where comparison results are not fully connected. Consequently, KRC becomes highly effective for estimating item ratings in a dynamic setting.

The structure of this paper is as follows. Section \ref{sec_dbt} outlines the problem setup and introduces the dynamic Bradley-Terry model. Section \ref{sec_method} elucidates the construction procedure of our proposed method. Section \ref{sec_theo} presents the theoretical properties of this method. Sections \ref{sec_simu} and \ref{sec_real} provide numerical results from simulations and the analysis of real-world data, illustrating our theoretical approach, respectively. Section \ref{sec_diss} provides a discussion and concluding remarks. The appendix contains detailed technical proofs.

\section{Dynamic Bradley-Terry model}
\label{sec_dbt}

Given $n$ items,  a sequence of binary comparison outcomes
$\{y_{ij}(t_k), t_k \in T_{ij}\}$ are available at timestamps 
$t_k \in T_{ij}$ for each pair of items $(i,j)$; $1\leqslant i<j \leqslant n$,
where $y_{ij}(t_k)=1$ if item $j$ is preferred over item $i$ at
at $t_k$ and $y_{ij}(t_k)=0$ otherwise, and $y_{ij}(t_k)+y_{ji}(t_k)=1$ for
$i \neq j$.  In practice, for instance, 
$y_{ij}=1$ could mean consumers prefer the item $j$ to $i$ or team $j$ beats team $i$ in 
sports tournaments.  For simplicity, we assume that the observation time $t_k$ is 
uniformly distributed over $[0,1]$ for any pair $(i,j)$. 

The dynamic Bradley-Terry model focuses on the probability of event $y_{i j}(t)=1$: 
\begin{assumption}[Dynamic Bradley-Terry model]\label{A_bt}
	The outcome probability $y_{ij}(t)$ is independent and follows the Bernoulli distribution;
	$1 \leq i<j \leq n$:
	\begin{equation}
		\mathbb{P}\left(y_{i j}(t)=1\right)=\frac{\pi_{j}(t)}{\pi_{i}(t)+\pi_{j}(t)}; 
		\quad \sum_{i=1}^{n}\pi_i(t)=1, 
	\end{equation}
	where $\bm{\pi}(t)=(\pi_1(t),\ldots,\pi_n(t))^\top$ refers to as a preference score at time $t$. 
\end{assumption}

\cite{Ludwig1994} initially applied the dynamic Bradley-Terry model to sports competition data analysis, integrating a state-space model to accommodate time-dependent features. Later studies primarily concentrated on the Bayesian framework. The dynamic Bradley-Terry model, supplemented by its Bayesian updates, plays a crucial role in the esteemed Glicko rating system \citep{Glickman1999, Glickman2001}. \cite{Cattelan2013} developed a clear-cut scheme that sequentially updates skill estimation by combining the previous skill level with the latest comparison results. Importantly, in contrast to the Bayesian methods, we refrain from making any prior assumptions or specifications about $\bm{\pi}(t)$ under Assumption \ref{A_bt}. This setup, as specified in Assumption \ref{A_bt}, aligns with the one put forward by \cite{MCHALE2011619} and \cite{bong2020nonparametric}. In \cite{MCHALE2011619} and \cite{bong2020nonparametric}, a score vector at a specific time point can be obtained by maximizing the time-weighted likelihood function. However, these methods are  computationally demanding by the need to solve nonlinear optimization problems.

\section{Kernel Rank Centrality}
\label{sec_method}

\subsection{Method}

To meet the computational challenges, we propose a spectral method aggregating pairwise comparisons over time with kernel smoothing. First, 
we construct an observation matrix 
$\hat{\bm{P}}(t)=[\hat{P}_{ij}(t)]_{1\leqslant i,j\leqslant n}$ for each outcome $\{y_{ij}(t)\}$ over time $t$:
\begin{equation}
	\label{matrix}
	\hat{P}_{ij}(t)=\left\{
	\begin{array}{ll}
		\displaystyle \frac{1}{n}\frac{\sum_{t_k\in T_{ij}} y_{ij}(t_k)K_h(t,t_k)}{\sum_{t_k\in T_{ij}}K_h(t,t_k)} & \text{if $i\neq j$,}   \\
		\displaystyle 1- \sum_{s\neq i} \hat{P}_{is}(t) & \text{if  $i = j$,}\\
	\end{array}
	\right. 
\end{equation}
where $K_h(t,t_k)=K(\frac{t-t_k}{h})$ is a kernel to be specified in practice and 
$h$ is a bandwidth. In \eqref{matrix}, $\hat{P}_{ij}(t)$ is locally smooth by $K$ over $t$. By
design, $\hat{\bm{P}}(t)$ is nonnegative with the row sum one by design, which 
defines a probability transition matrix of a Markov chain.

The motivation for our method originates from multiple-round voting with $n$ candidates \citep{langville2012s}, where we are interested in estimating the percentage of votes for each candidate at different time points. In each round of voting, candidate $i$ receives a certain percentage of votes from candidate $j$ if candidate $i$ defeats candidate $j$ at that time-point. Assume that pairwise contests are close to the time $t$, with a high weight assigned by the kernel. This assumption allows us to approximate the probability transition matrix $\bm{P}^*(t)= [P^*_{ij}(t)]_{1\leqslant i,j\leqslant n}$ of a reversible Markov chain by the matrix $\hat{\bm{P}}(t)=[\hat{P}_{ij}(t)]_{1\leqslant i,j\leqslant n}$ when $\frac{\sum_{t_k\in T_{ij}} y_{ij}(t_k)K_h(t,t_k)}{\sum_{t_k\in T_{ij}}K_h(t,t_k)} 
\approx \frac{\sum_{t_k\in T_{ij}} y^*_{ij}(t_k)K_h(t,t_k)}{\sum_{t_k\in T_{ij}}K_h(t,t_k)}\approx\frac{\pi_j(t)}{\pi_i(t)+\pi_j(t)}$:
\begin{equation}\label{f_P*}
	P^*_{ij}(t)=\left\{
	\begin{array}{ll}
		\displaystyle \frac{1}{n}\frac{\pi_j(t)}{\pi_i(t)+\pi_j(t)} & \text{if $i\neq j$,}   \\
		\displaystyle 1- \sum_{s\neq i} \frac{1}{n} \frac{\pi_s(t)}{\pi_s(t)+\pi_i(t)} & \text{if  $i = j$,}
	\end{array}
	\right.
\end{equation}
where $\bm{\pi}(t)$ denotes the preference score vector in Assumption \ref{A_bt}
and $y^*_{ij}(t_k)=\mathbb{E}(y_{ij}(t_k))=\frac{\pi_j(t_k)}{\pi_i(t_k)+\pi_j(t_k)}$ represents the probability that candidate $j$ beats candidate $i$ at time $t_k$.
By Corollary 2.4.15 in \cite{bremaud2013markov}, the stationary distribution of a Markov chain is exactly the normalized latent skill vector if  $\pi_i(t)P^*_{ij}(t)=\pi_j(t)P^*_{ji}(t)$.

To estimate $\bm{\pi}(t)$, we use the dynamic Bradley-Terry model assumption that the ratio of the percentage of votes between two candidates approximates the ratio of their skill levels. Specifically, we assume that $\frac{\pi_j(t)}{\pi_i(t)+\pi_j(t)}$ is approximately equal to $\frac{\sum_{t_k\in T_{ij}} y_{ij}(t_k)K_h(t,t_k)}{\sum_{t_k\in T_{ij}}K_h(t,t_k)}$, where $y_{ij}(t_k)$ represents the outcome of the pairwise contest between candidates $i$ and $j$ at time $t_k$, and $T_{ij}$ represents the set of time points when candidate $i$ competes against candidate $j$. The kernel function $K_h(t,t_k)$ assigns a weight to each pairwise contest based on its proximity to time $t$ and the kernel bandwidth $h$. By solving this equation for $\bm{\pi}(t)$, we obtain an estimated $\hat{\bm{\pi}}(t)$ of votes for each candidate at time $t$.

The above discussion leads to the proposed method, which computes the stationary distribution of the 
Markov chain defined by $\hat{\bm{P}}(t)$:
\begin{equation}
	\left\{
	\begin{array}{ll}
		\displaystyle \hat{\bm{\pi}}(t)^\top=\hat{\bm{\pi}}(t)^\top\hat{\bm{P}}(t),\\
		\displaystyle \sum_{i=1}^n\hat{\pi}_i(t)=1.
	\end{array}
	\right.
	\label{eigen}
\end{equation}
The solution of \eqref{eigen}, denoted by $\hat{\bm{\pi}}(t)$,  is unique under the assumption that the Markov chain defined by $\hat{\bm{P}}(t)$ is irreducible in that $\hat{\bm{P}}(t)$ is 
positive definite for any $t$. Technically, to ensure irreducibility, we replace
$\hat{\bm{P}}(t)$ by its regularized version $[(1-\sigma_n)\hat{P}_{ij}(t)+\frac{\sigma_n}{n}]_{1\leqslant i,j\leqslant n}$, where $\frac{\sigma_n}{n}>0$ is a small random teleportation probability \citep{Gleich2015}, e.g., $\frac{\sigma_n}{n}=\frac{1}{n^2}$, although
Lemma 2 says that unregularized $\hat{\bm{\pi}}(t)$ yields
irreducibility with a high probability.

Then, $\hat{\bm{\pi}}(t)$ ranks items, which is called Kernel Rank Centrality (KRC) subsequently.

Note that, by using an appropriate kernel function, both the static Rank Centrality \citep{Negahban2017} and the nearest neighbor-based Rank Centrality \citep{karle2021dynamic} are two special cases of our method.

Algorithm \ref{Algorithm1} summarizes the computational strategy for the proposed method.
\begin{algorithm}
	\caption{Kernel Rank Centrality}
	\begin{algorithmic}[1]
		\label{Algorithm1}
		\REQUIRE Binary comparisons $\{y_{ij}(t_k)\}$, kernel, time $t$, and bandwidth $h$.
		\ENSURE The score vector $\hat{\bm{\pi}}(t)$.
		\STATE
		Define the stochastic matrix $\hat{\bm{P}}(t)$ by (\ref{matrix});
		\STATE
		Compute the solution $\hat{\bm{\pi}}(t)$ of (\ref{eigen});
		\RETURN 
		$\hat{\bm{\pi}}(t)$.
	\end{algorithmic}
\end{algorithm}

\subsection{Advantages of Kernel Rank Centrality}

\begin {figure*}[!t]
\centering
\renewcommand{\dblfloatpagefraction}{.95}
\includegraphics[width=0.95\textwidth]{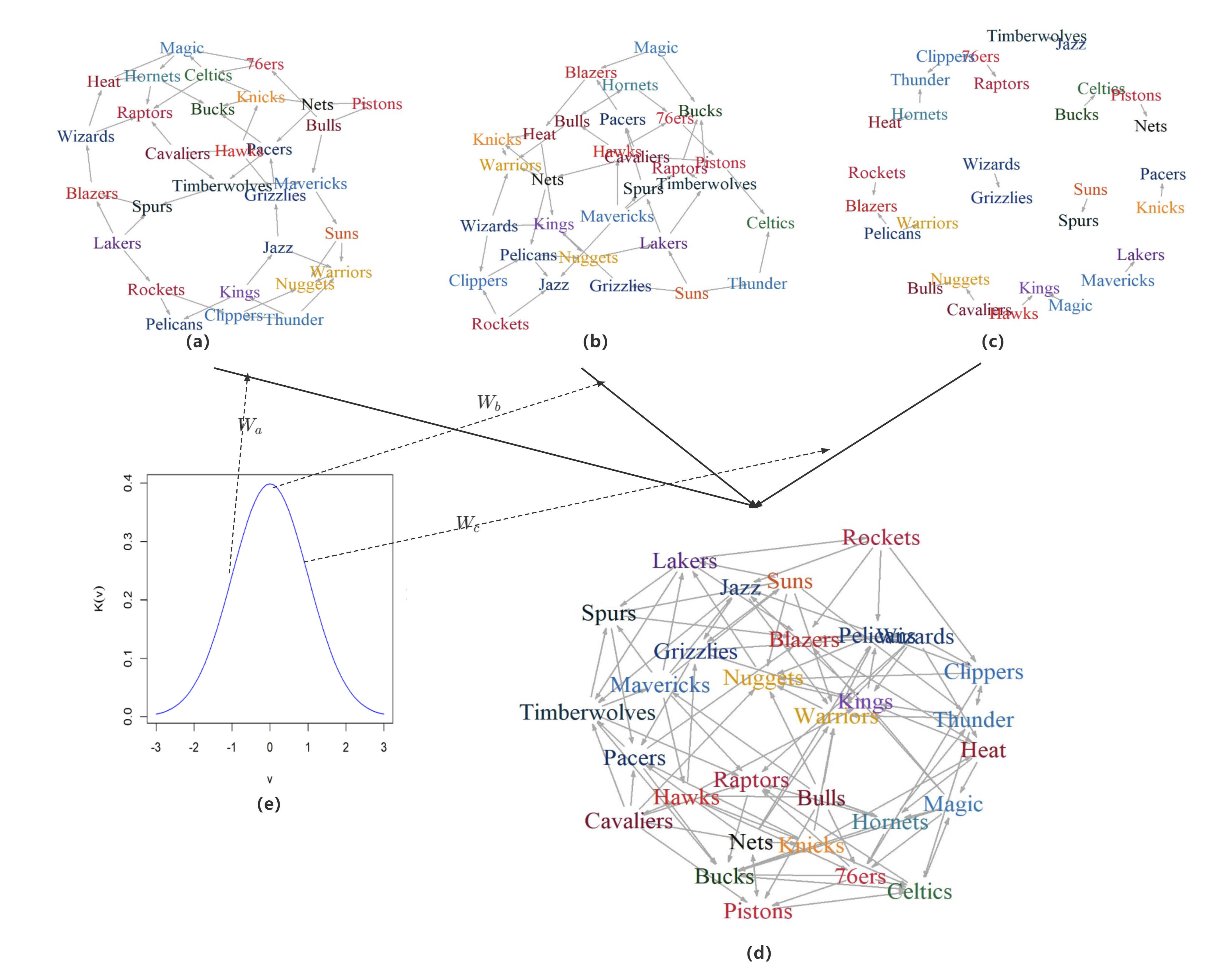}
\caption{Aggregation of the NBA game graphs. Figures (a)-(c) display the NBA games results in the first week
	of the 2018-2019 season, the second week of the 2018-2019 season, and the first three days of the third week of
	the 2018-2019 season. Figure (d) presents the aggregation graph via all the games with time-dependent weights
	defined by the Gaussian kernel function, which is displayed in Figure (e).\label{fig_example}}
\end{figure*}

This section presents an NBA data example to demonstrate the advantages of
the proposed dynamic method over the original static Rank Centrality
methods of \cite{Negahban2017} in Figure \ref{fig_example}. 
In the NBA data example presented in Figure \ref{fig_example}, sub-figures (a)--(c) show directed graphs generated by comparison results in three periods. The arrows in each graph represents the strength score transition from one node to another in the Markov chain. One common approach for treating dynamic ranking is to estimate the score vector $\pi(t)$ independently for each graph, as adopted in the nearest neighbor frame \citep{karle2021dynamic}. However, this method may not provide smoothly varying estimates and may suffer from insufficient comparisons for graph connectivity when the neighborhood is too narrow, such as a week of NBA game data can not ensure the connectivity. The proposed KRC method overcomes these issues by using a Gaussian kernel to generate time-dependent weights and aggregating (a)--(c) into a connected graph (d), ensuring the connectivity automatically. Compared to the static Rank Centrality method \citep{Negahban2017}, the KRC method can estimate strength curves over a given period and has the advantage of exact online updating,  which improves computation efficiency.

As pointed out by \cite{Negahban2017} and  \cite{Jain2020}, the computational cost of maximum likelihood estimation (MLE) for the Bradley-Terry model is much higher compared to Rank Centrality. In a dynamic setting, this computational issue becomes further complex in the presence of time-varying comparisons. However, an advantage of the proposed KRC method is the availability of exact online updating, which enables us to update \eqref{eigen} with a few matrix operations rather than optimization. Online updating significantly improves the computation efficiency of the method.

\subsection{On-line Ranking}

The proposed ranker \eqref{eigen} has an online version that updates the estimate $\hat{\bm{\pi}}(t)$ as new observations for some pair $(i,j)$ come in. Specifically, we sequentially apply the Sherman–Morrison rank-one updating formula \citep{update2006}. This formula updates a stationary vector when a transition matrix perturbs only one row.

\begin{lemma}[Rank-One Updating]
Let $\bm{P}_{old}$ be a transition matrix.
If the ith row $\bm{P}_{i.,old}$ of $\bm{P}_{old}$ is updated to produce $\bm{P}_{i.,new} = \bm{P}_{i.,old} -\bm{\delta}^\top$,
the ith row of $\bm{P}_{new}$, and if $\bm{\pi}_{old}$ and $\bm{\pi}_{new}$ are the respective  stationary vectors of $\bm{P}_{old}$ and $\bm{P}_{new}$, then $\bm{\pi}_{new}=\bm{\pi}_{old}-\bm{\phi}$, where
\begin{equation}
	\bm{\phi}^\top=\left[\frac{{\pi}_{i,old}}{1+\bm{\delta}^\top\bm{A}^{\#}_{.i,old}}\right]\bm{\delta}^\top\bm{A}^{\#}_{old}
	. 
	\label{f_updatePI}
\end{equation}
$\bm{A}^{\#}_{old}$ is the group inverse of $\bm{A}_{old}=\bm{I}-\bm{P}_{old}$. Multiple row updates to $\bm{P}_{old}$ requires applying this formula one
row at a time and the group inverse must be sequentially updated. The updating formula
for $\bm{A}^{\#}_{new}=(\bm{I}-\bm{P}_{new})^{\#}$ is as follows:
\begin{equation} 
	\bm{A}_{new}^{\#}=\bm{A}_{old}^{\#}+\bm{e}\bm{\phi}^\top(\bm{A}_{old}^{\#}-\frac{\bm{\phi}^\top A_{.i,old}^{\#}}{\pi_{i,old}}\bm{I})-\frac{A_{.i,old}^{\#}\bm{\phi}^\top}{\pi_{i,old}}.\label{f_updateA}
\end{equation} 
\end{lemma}

The updating formula mainly relies on the group inverse matrix, $\bm{A}_{old}^{\#}$. The group inverse of a square matrix $\bm{A}$, which is a special case of the Drazin inverse, is the unique matrix $\bm{A}^{\#}$ satisfying $\bm{A}\bm{A}^{\#}\bm{A}=\bm{A}$, $\bm{A}^{\#}\bm{A}\bm{A}^{\#}=\bm{A}^{\#}$ and $\bm{A}^{\#}\bm{A}=\bm{A}\bm{A}^{\#}$.
To perform updating, we define the rank-one updating operator $\mathcal{ROU}:{\bm{\pi}_{old},\bm{A}^{\#}_{old},\bm{\delta},i}\rightarrow{\bm{\pi}_{new},\bm{A}^{\#}_{new}}$ by combining the updating formulas.

For the KRC updating, when a new observation for the pair $(i,j)$ arrives at time $t_{new}$, the change only affects the $i$-th and $j$-th rows of the transition matrix. Let $\hat{\bm{P}}_{new}(t)$ and $\hat{\bm{P}}_{old}(t)$ denote the updated and old transition matrices, respectively. The perturbation vector for the $i$-th row, $\bm\delta^{(i)}$, can be defined as:
\begin{equation}
{\delta}_{k}^{(i)}=\left\{
\begin{array}{ll}
	\hat{P}_{ij,old}(t)-\hat{P}_{ij,new}(t) & \text{if $k=j$},\\
	-(\hat{P}_{ij,old}(t)-\hat{P}_{ij,new}(t)) & \text{if $k=i$},\\
	0& \text{otherwise.}
\end{array}
\right.
\end{equation}
Due to the construction of $\hat{\bm{P}}(t)$, the perturbation vector for the $j$-th row, $\bm\delta^{(j)}=-\bm\delta^{(i)}$. Thus, we update the transition matrix in the $i$-th and $j$-th rows and then perform rank-one updating \citep{update2006} twice to solve equation \eqref{eigen} online.

We present the detailed online updating method in Algorithm \ref{Algorithm2}. This updating algorithm for the KRC method can be completed in just a few steps, avoiding the iterative optimization process in the MLE-based method.
\begin{algorithm}[h]
\caption{Online updating}
\begin{algorithmic}[1]
	\label{Algorithm2}
	\REQUIRE The new comparison observation $y_{ij}(t_{new})$, the kernel function, the time point $t$, bandwidth parameter $h$, old transition matrix $\hat{\bm{P}}_{old}(t)$, old group inverse matrix $\bm{A}_{old}^{\#}(t)$, old estimation $\hat{\bm{\pi}}_{old}(t)$.
	
	\ENSURE The new score vector $\hat{\bm{\pi}}_{new}(t)$ and the new group inverse matrix $\bm{A}_{new}^{\#}(t)$.
	
	\STATE	
	
	Compute the new transition probability $\hat{P}_{ij,new}(t)$.
	\STATE
	Compute the difference vectors $\bm\delta^{(i)}$ and $\bm\delta^{(j)}$.
	
	\STATE 
	Update 
	
	$\{\hat{\bm{\pi}}_{tmp}(t),\bm{A}_{tmp}^{\#}(t)\}=\mathcal{ROU}(\hat{\bm{\pi}}_{old}(t),\bm{A}_{old}^{\#}(t),\bm\delta^{(i)},i)$.

	\STATE		
	Update 
	
	$\{\hat{\bm{\pi}}_{new}(t),\bm{A}_{new}^{\#}(t)\}=\mathcal{ROU}(\hat{\bm{\pi}}_{tmp}(t),\bm{A}_{tmp}^{\#}(t),\bm\delta^{(j)},j)$.
	
	\RETURN 
	$\hat{\bm{\pi}}_{new}(t)$ and $\bm{A}_{new}^{\#}(t)$.
\end{algorithmic}
\end{algorithm}

\section{Learning Theory}
\label{sec_theo}

This section investigates various theoretical issues for the proposed ranker,
including the uniqueness of \eqref{eigen}, the rates of convergence, and 
most importantly, its asymptotic distribution. 

The following conditions are assumed.

\begin{assumption}[Capability ratios]\label{A_pi}
Assume that $\sup_{t\in[0,1]} \frac{\max_{i}\pi_i(t)}{\min_{i}\pi_i(t)}\leqslant \kappa$. Also, 
$\pi_i(t)$ is three times continuously differentiable; $1\leqslant i
\leqslant n$.
\end{assumption}

Assumption \ref{A_pi} requires that the maximal discrepancy of the item's skill score ratio 
is upper bounded and $\pi_i(t)$ is sufficiently smooth as in 
\cite{li2007nonparametric}. 

\begin{assumption}[Kernel]\label{A_kf}
Kernel $K(\cdot,\cdot)$ is nonnegative and symmetric, satisfying 
\begin{align*}
	&\int_{-\infty}^{+\infty}K(v) \dif v=1, \int_{-\infty}^{+\infty}v^2K(v) \dif v < \infty.
\end{align*}
\end{assumption}

Let $M=\min_{i,j} \#|T_{ij}|$.

\begin{lemma}[Uniqueness]\label{lemma_unique}
Under Assumptions \ref{A_bt} and \ref{A_pi}, when $n,M 
\rightarrow \infty$, the solution of \eqref{eigen} exists and is
unique with probability at least $1-O((nM)^{-10})$.
\end{lemma}

Lemma \ref{lemma_unique} says that the solution of \eqref{eigen} is unique if there
are abundant observations. Next we provide an $\ell_{2}$ 
error bound in terms of  $n$, $M$, and $h$.

The next theorem establishes the convergence rate for the KRC method.

\begin{theorem}[Rates of convergence]\label{Theorem_1}
Assume that Assumptions \ref{A_bt}--\ref{A_kf} are met. Then, for some constants $C_1,C_2>0$ and 
any $t\in (0,1)$, as $n \rightarrow \infty$ and $Mh\rightarrow \infty$,  
$$\mathbb{P}\left(\frac{\left\|\hat{\bm{\pi}}(t)-\bm{\pi}(t)\right\| }{\left\|\bm{\pi}(t)\right\|}\leqslant  C_1\sqrt{\frac{1}{nMh}}+C_2h^2 \right)\rightarrow 1,  
$$
provided that the bandwith $h \rightarrow 0$. For the entrywise error, under 
an additional assumption that $nMh^7\rightarrow 0$, there exist positive constants $C_3$ and $C_4$ such that, for any $t\in (0,1)$,
$$
\mathbb{P}\left(\frac{\|\hat{\bm{\pi}}(t)-\bm{\pi}(t)\|_{\infty}}
{\|\bm{\pi}(t)\|_{\infty}}\leqslant C_3\sqrt{\frac{\log(nM)}{nMh}}+C_4{h^2}\right)\rightarrow 1.
$$

\end{theorem}
The first term $\sqrt{\frac{1}{nMh}}$ is attributed to the variance of binary comparison, while the second term represents the bias due to the dynamic skill. Interestingly, the bias does not increase with an increased item number $n$. Importantly, this error bound suggests that the optimal bandwidth is of order $(nM)^{-\frac{1}{5}}$, which achieves the optimal trade-off between the bias and variance in that $\sqrt{\frac{1}{nMh}} \approx h^2$.

Next, we will present the asymptotic normality for the KRC method.

\begin{theorem}[Finite-dimensional asymptotic normality]\label{theorem_clt}
Assume that Assumptions \ref{A_bt}-\ref{A_kf} are met. Then, $\hat{\bm{\pi}}(t)$ is asymptotically normal for any subvector of a fixed length 
as $Mh\rightarrow \infty$ and $n \rightarrow \infty$ provided that 
$nMh^7\rightarrow 0$, that is, for any fixed $s\in [n]$ and $t\in (0,1)$, 
$$
\left(\begin{array}{cc}
	\alpha_1(t)(\hat{\pi}_1(t)-\pi_1(t)-\beta_1(t) h^2)&  \\
	\vdots&
	\\
	\alpha_s(t)(\hat{\pi}_s(t)-\pi_s(t)-\beta_s(t) h^2)& 
\end{array}\right)
\buildrel d\over\longrightarrow N(\bm{0},\bm{I_s}),
$$
where
\begin{align*}
\alpha_i(t)=\sqrt{\frac{(\sum_{j\neq i}y_{ij}^*(t))^2}{\sum_{j\neq i}{\frac{1}{M_{ij}h}(\pi_i(t)+\pi_j(t))^2 y^*_{ij}(t)(1-y^*_{ij}(t))\int K^2(v)\mathrm{d}v}}},
\end{align*}
\begin{align*}
\beta_i(t)=\sum_{1\leqslant k<l\leqslant n}(A^{\#}_{li}(t)-A^{\#}_{ki}(t))\frac{\pi_k(t)+\pi_l(t)}{n}\ddot y^*_{kl}(t)\int v^2K(v) \mathrm{d}v,
\end{align*}
$\ddot y^*_{kl}(t)$ is the second order derivative and $\bm{A}^{\#}(t)$ is the group inverse of $\bm{A}(t)=\bm{I}-\bm{P}^*(t)$.
\end{theorem}

The covariance matrix in the asymptotic normal distribution is diagonal, which means that the dependency from the comparisons between pairs is absent with the diverging $n$. The independence between $\hat{\bm{\pi}}(t)$ is consistent with the finding in the MLE of the static Bradley-Terry model \citep{Simons1999}. The extra condition $nMh^7\rightarrow 0$ ensures that the remaining bias term, which is of high order $O(\frac{h^3}{n})$, is much smaller than the variance term.
The $\ell_{2}$ error analysis uses an inequality induced by the spectral norm of the error matrix defined by $\bm{E}(t)=\hat{\bm{P}}(t)-\bm{P}^*(t)$. For the $\ell_{\infty}$ error and the asymptotic distribution theory, we conduct entrywise analysis through a novel tool of the group-inverse matrix and its approximation. The group inverse is often involved in questions concerning Markov chains. Using an appropriate approximation technique, we give a sharp expansion for the Rank Centrality type estimator. The asymptotic normality and $\ell_{\infty}$ convergence rate are direct applications of this expansion. We will describe this novel tool and the expansion results in detail in the Appendix.

It is important to note that all the properties obtained for the spectral method when $n\rightarrow \infty$ continue to hold without the condition $Mh\rightarrow\infty$. This condition, together with the assumption on $y^*_{ij}(t)$, allows us to bound the bias term as the order $O(h^2)$. When there are not enough longitudinal observations but the error of $\|\bm{E}(t)\|$ is still guaranteed to be less than 
$1-\lambda_{max}(\bm{P}^*(t))$, the proof strategy still applies except that the size of the bias term increases. For example, when $Mh$ is a constant, which means we have a finite number of observations for $(i,j)$ in an interval with length $h$, the bias term would increase to $O(\frac{h}{n})$, but the variance part can still be analyzed using the same argumentation.

\section{Simulations}
\label{sec_simu}

This section presents simulation results to investigate the operating characteristics of the KRC method. 
In simulations, we simulate as follows:
\begin{enumerate}
\item Generate $\bm{\alpha}=[\alpha_i]_{1\leqslant i\leqslant n}$ from the uniform distribution U(1,3).
\item For $1\leqslant i< j\leqslant n$ and $1\leqslant m \leqslant M$, sample $t_m$ from U(0,1) and compute the capability
parameter ${\pi}_i(t_m)=\alpha_i+sin(5\alpha_i t_m)$. Then generate the comparison result $y_{ij}(t_m)$ from $\mathrm{Bernoulli}(\frac{\pi_j(t_m)}{\pi_i(t_m)+\pi_j(t_m)})$.
\end{enumerate}

In the first experiment, we increase $n$ from 10 to 100 by 10 to see the impact of the number of items on the performance
while setting $M=50,100,150,200$ to see the effect of the sample size. For evaluation metric,
we use the average normalized root mean squared error (RMSE), i.e. average $\ell_2$ error defined as $\frac{1}{M-1}\sum_{k=1}^{M-1}\frac{\left\|\hat{\pi}(t_k)-\pi(t_k)\right\| }{\left\|\pi(t_k)\right\|},t_k=\frac{k}{M},$ and the max $\ell_{\infty}$ error defined as 
$\max_{1\leqslant k\leqslant M-1}\frac{\left\|\hat{\pi}(t_k)-\pi(t_k)\right\|_{\infty} }{\left\|\pi(t_k)\right\|_{\infty}},t_k=\frac{k}{M}$,
based on 100 simulation replications. Moreover, we use the Gaussian kernel $K(x)=\frac{1}{\sqrt{2\pi}}\exp{(-\frac{x^2}{2}})$.  And for the first experiment, we set the fixed bandwidth as $h=0.1$.

\begin {figure}[h]
\centering
\includegraphics[width=0.9\textwidth]{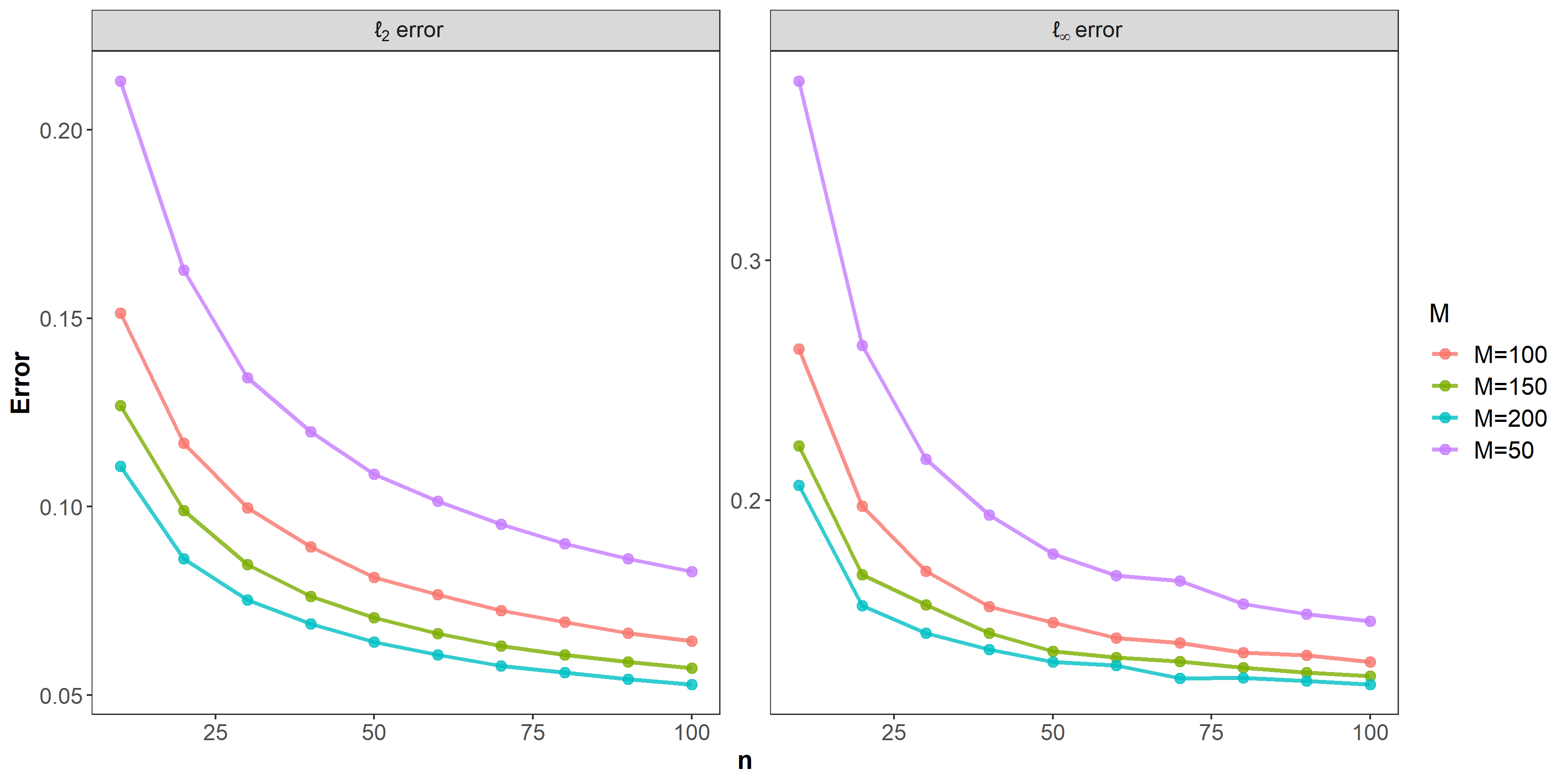}
\caption{Estimation error as a function of $n$ for different values of $M$.\label{fig_mn} }
\end{figure}

As shown in Figure \ref{fig_mn}, the estimation error decreases as $M$ increases or $n$ increases. This observation
is in accordant with the result of Theorem \ref{Theorem_1}.

\begin {figure}[!t]
\centering
\includegraphics[width=0.8\textwidth]{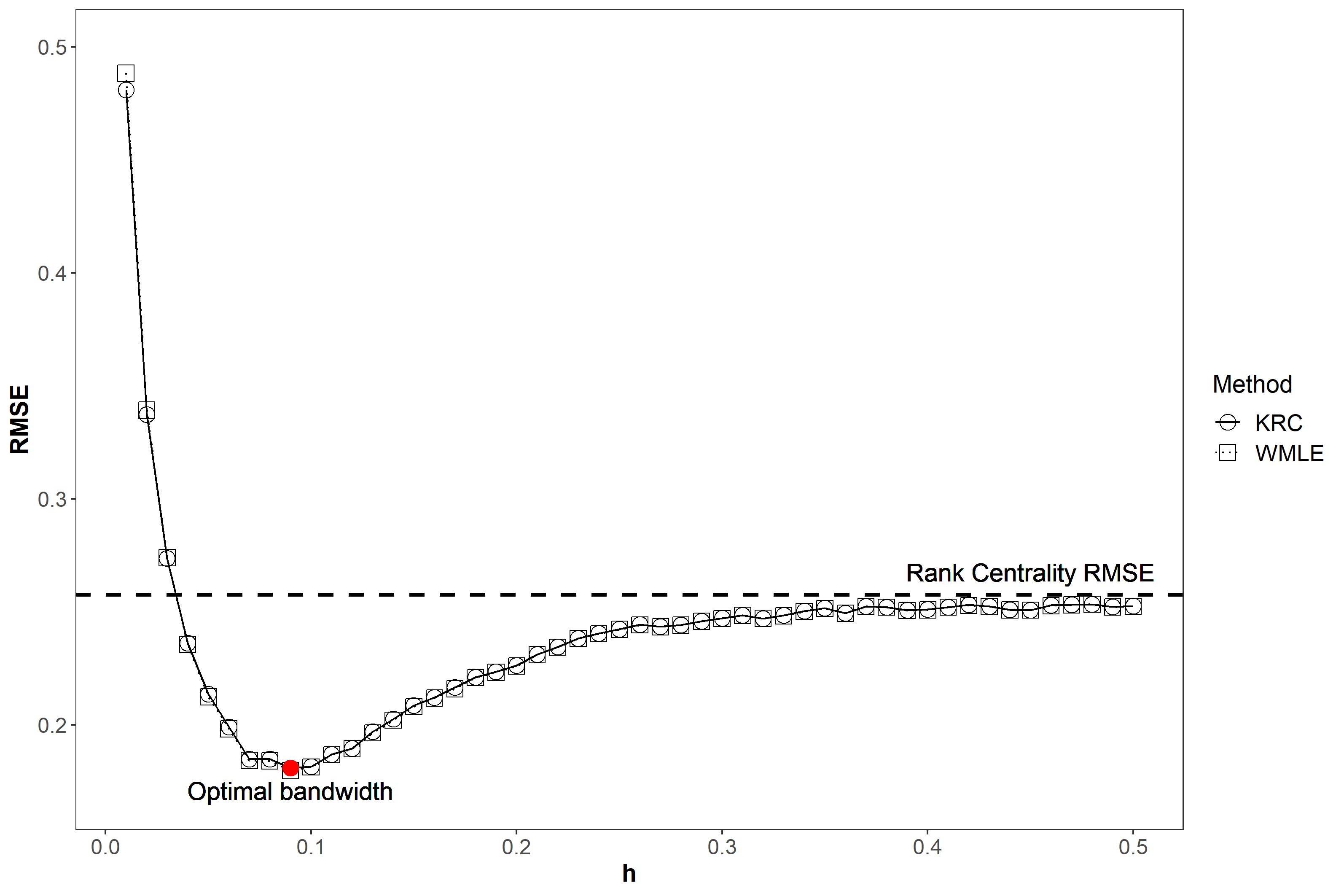}
\caption{RMSEs of Rank Centrality, Kernel Rank Centrality, and WMLE. The red point represents the optimal bandwidth with the minimum RMSE, while circles and squares represent KRC and WMLE, respectively.\label{fig_band}}
\end{figure}

Concerning the bandwidth choice on the estimation of KRC, we draw the RMSEs under different bandwidths when $n=10$ and $M=50$ in Figure \ref{fig_band}. As shown, the estimation error decreases initially, then increases, and eventually reaching equilibrium at the error of the Rank Centrality, as the bandwidth increases. 
The observation is consistent with the theoretical result of Theorem \ref{Theorem_1}. 

Next, we also compare KRC with the weighted maximum likelihood estimate (WMLE) of \cite{bong2020nonparametric} under various bandwidths in Figure \ref{fig_band}, where WMLE is
obtained by
\begin{equation}
\hat{\bm{\pi}}(t)=\arg\min_{\begin{array}{c}
	\sum_i\pi_i=1\\
	\bm{\pi}>0
	\end{array}
	}\sum_{i\neq j} \frac{\sum_{t_k\in T_{ij}} y_{ij}(t_k)K_h(t,t_k)\log\frac{\pi_{j}}{\pi_{i}+\pi_{j}}}{\sum_{t_k\in T_{ij}}K_h(t,t_k)},
\end{equation}
In the experiment, we solve the nonlinearly constrained optimization with the MM algorithm for the Bradley-Terry model 
\citep{Hunter2004}.

\begin {figure}[!t]
\centering
\includegraphics[width=0.8\textwidth]{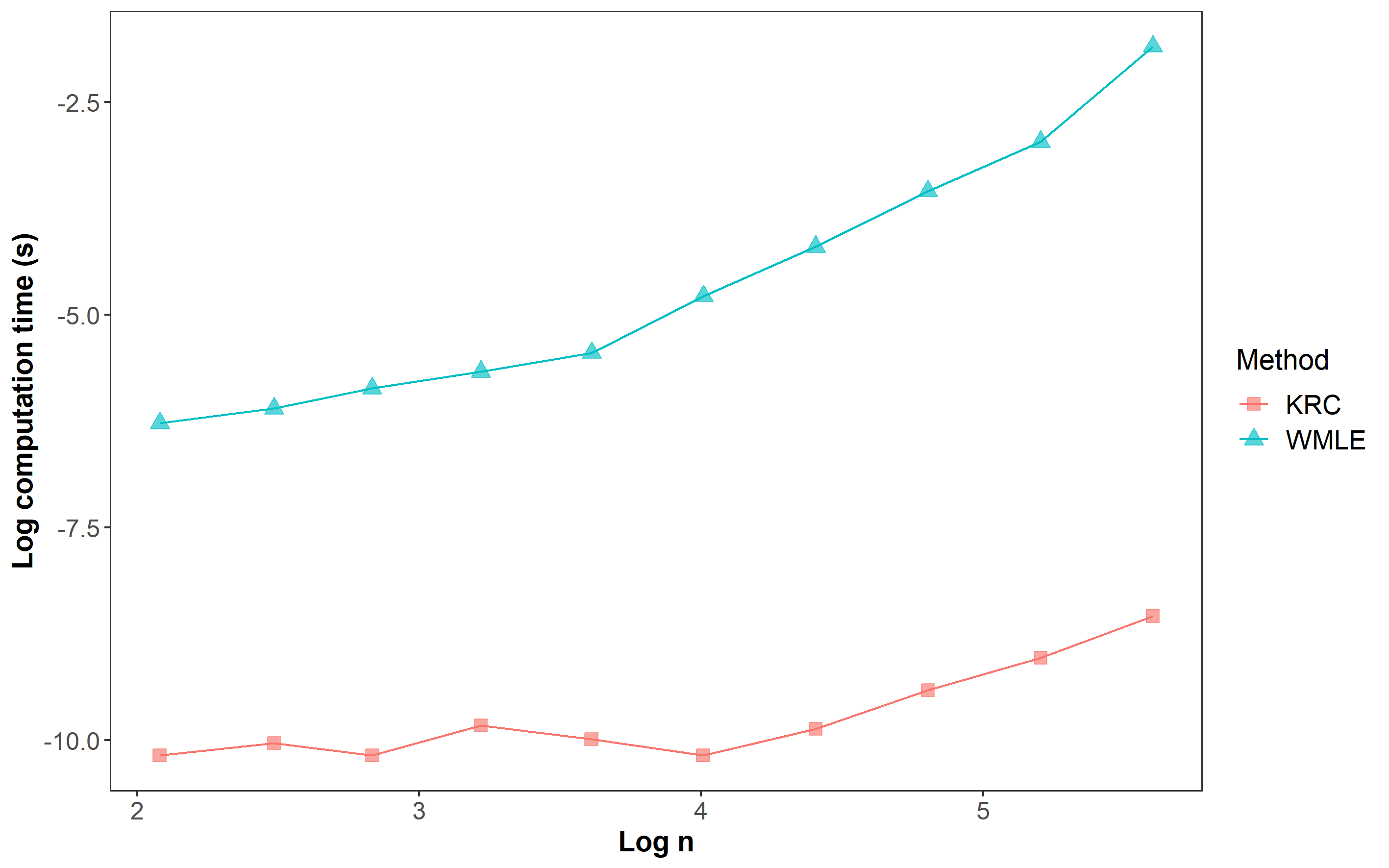}
\caption{Log-log plot of the time taken to execute KRC and WMLE.\label{fig_time}}
\end{figure}

As demonstrated in Figure \ref{fig_band}, the error achieved by our Kernel Rank Centrality method is nearly indistinguishable from that of WMLE. To further illustrate the advantage of the KRC method, we compared its computational efficiency with that of the WMLE under different problem sizes in Figure \ref{fig_time}. Although the stationary distribution of the Markov chain can be computed using linear algebraic techniques, we used the power method in our experiment. The results show that the computational cost of the KRC method is significantly lower than that of the WMLE. Additionally, the advantage in computational efficiency becomes more pronounced as $n$ increases.

\begin {figure}[!t]
\centering
\includegraphics[width=0.8\textwidth]{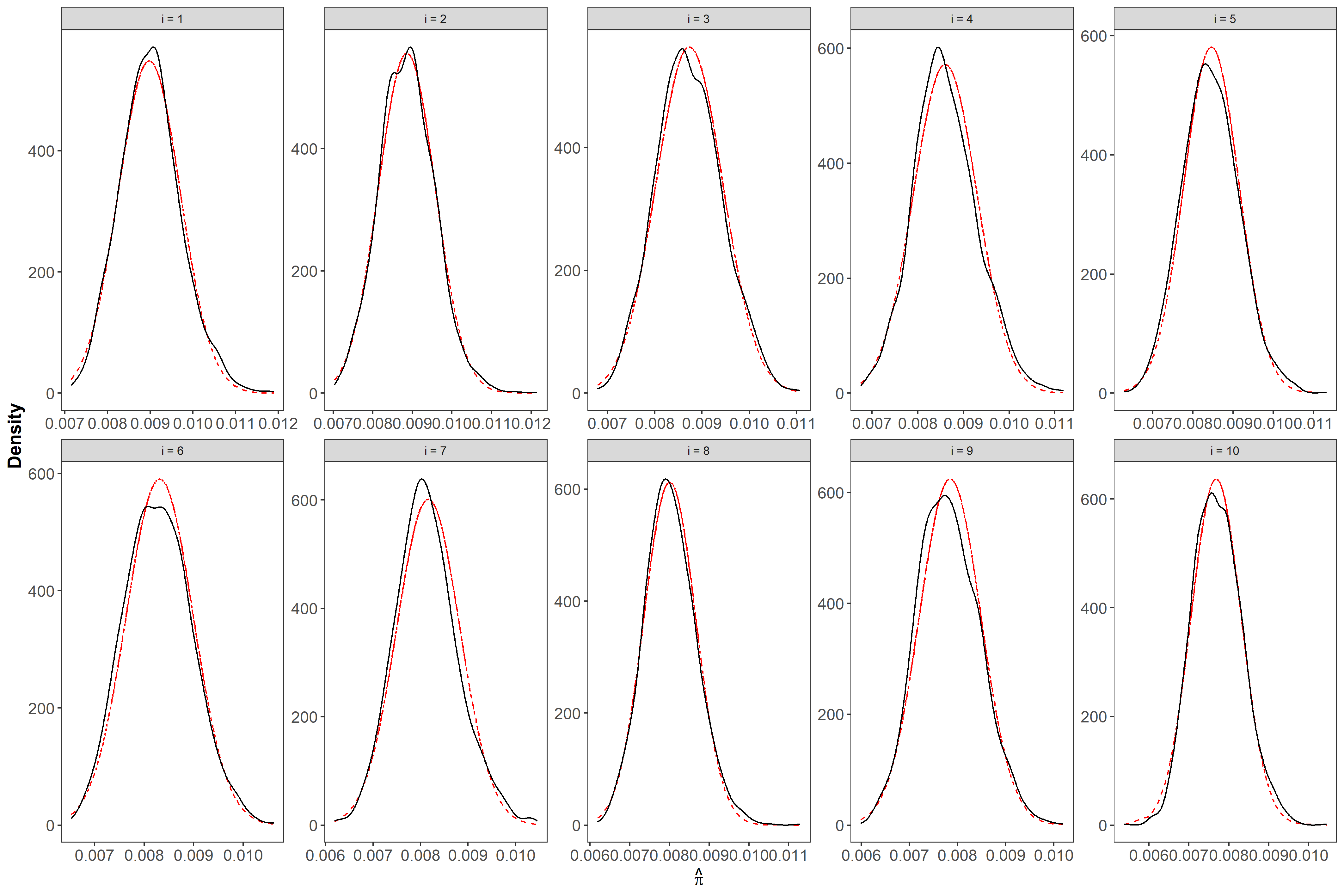}
\caption{The distribution of $\hat{\bm{\pi}}$. The black solid line represents the empirical distribution and the red dashed line represents the normal distribution in theory.\label{fig_dis}}
\end{figure}

\begin {figure}[!t]
\centering
\includegraphics[width=0.8\textwidth]{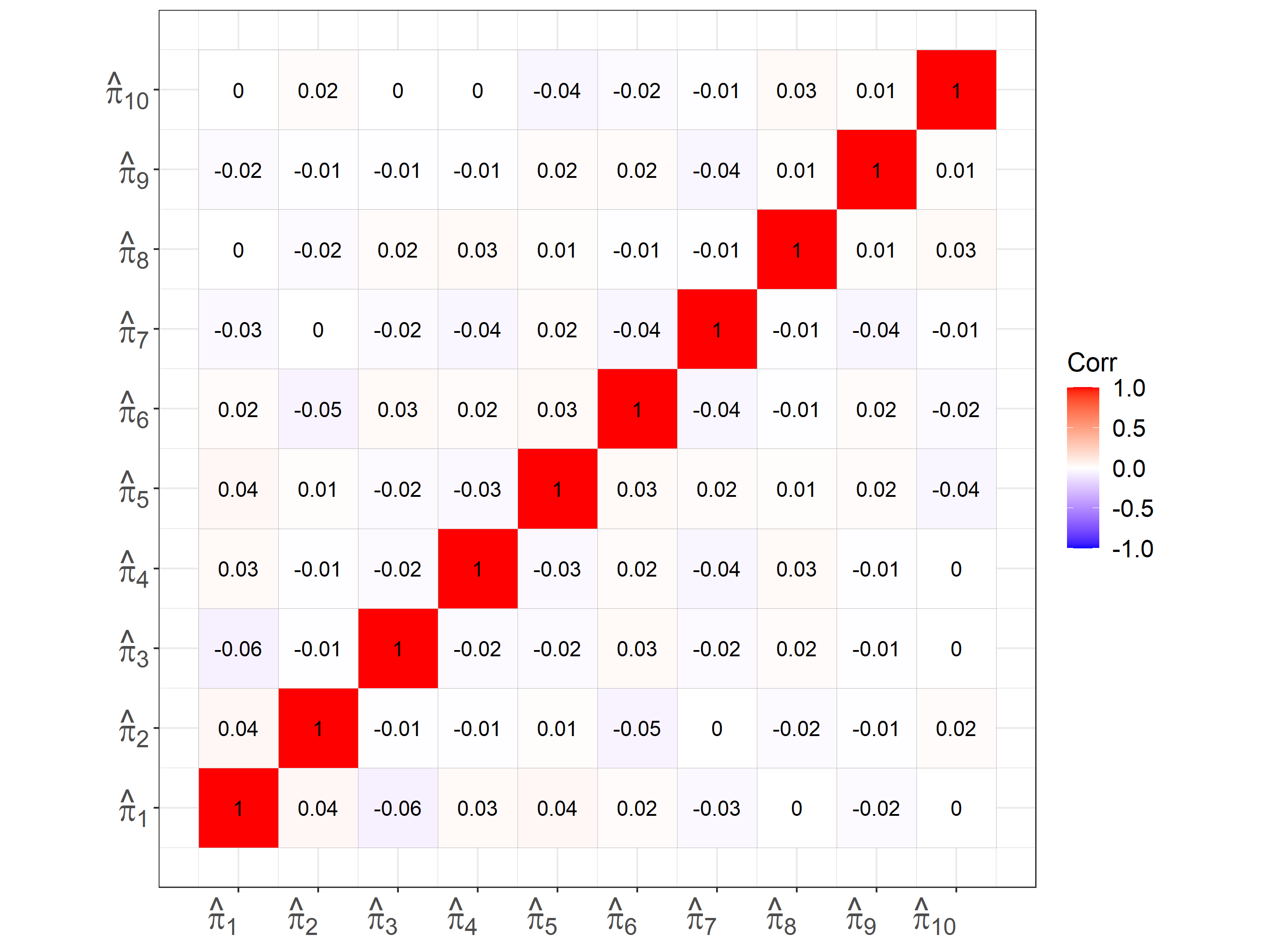}
\caption{Correlation matrix of $\hat{\bm{\pi}}$. \label{fig_corr}}
\end{figure}

To illustrate the asymptotic distribution theory, we simulated the distribution of $\hat{\bm{\pi}}$ with $n=100$ and $m=200$ for this experiment. To reduce the bias, we used an under-smoothing approach \citep{chen2017} with a smaller bandwidth $h=0.01$. By comparing the empirical distribution of $(\hat{{\pi}}_1(0.5),\hat{{\pi}}_2(0.5),\cdots,\hat{{\pi}}_{10}(0.5))^{\top}$ with its theoretical distribution defined in Section \ref{sec_theo} in Figure \ref{fig_dis}, we see that the empirical density curve closely resembles the density of the normal distribution, which aligns with the marginal distribution according to the theory. To examine the diagonal elements of the covariance matrix, we included the correlation heatmap in Figure \ref{fig_corr}. The correlation coefficient of any pair is indeed close to zero, which is consistent with the asymptotic independence result suggested by Theorem \ref{theorem_clt}.

\section{Real Data Analysis}
\label{sec_real}
To test our model in practical settings, we consider dynamically estimating the abilities of National Basketball Association (NBA) teams over multiple seasons. We collected NBA data for ten regular seasons, from 2009-2010 to 2018-2019, from www.nba.com. During each NBA season,  each team plays 82 regular games, resulting in 1230 regular season games. First, we test the forecasting performance of the Kernel Rank Centrality method using the following strategy:

\begin{enumerate}
\item Encode the time of the $k$-th game day in the $l$-th season as $l-1+\frac{k}{N_l+1}$, where $N_l$ is the size of the set of game days in the $l$-th season.
\item Take the games of the first seven seasons as the base and compute the team capability vector for each time point $t$ in the left three seasons with the data before $t$.
\item For each game in the left three seasons, we predict the winner of the game as the team with a larger skill score.
\end{enumerate}

 The proportion of correctly predicted matches, as the predictive accuracy, is used to measure performance. To determine whether the results produced by our model are reasonable, we compare our method with the relevant, openly available NBA Elo ratings \citep{silver2015we}. The Elo rating system is one of the most popular methods for estimating the competitors' abilities in sports \citep{KOVALCHIK20201329}. The Elo rating is a dynamic system that updates the rating after each game. We compare the predicting performance of the Elo rating with our method under different bandwidths; see Table \ref{tab1}.

\begin{table}[h]
\caption{Forecasting accuracy. \label{tab1}}
\centering
\begin{tabular}{lllll}
\hline
& 2016-2017                     & 2017-2018                     & 2018-2019                     & Total                         \\ \hline
RC          & 0.6114                        & 0.5732                        & 0.5935                        & 0.5927                        \\
KRC(h=0.1)  & 0.6228                        & {\color{red} 0.6333} & 0.6341                        & 0.6301                        \\
KRC(h=0.5)  & 0.6301                        & 0.6317                        & 0.6407                        & 0.6341                        \\
KRC(h=1)    & {\color{red} 0.6455} & 0.6325                        & 0.6366                        & {\color{red} 0.6382} \\
KRC(h=1.5)  & 0.6301                        & 0.6081                        & 0.6138                        & 0.6173                        \\
MLE         & 0.6179                        & 0.5740                        & 0.5919                        & 0.5946                        \\
WMLE(h=0.1) & 0.6089                        & 0.6260                        & 0.6358                        & 0.6236                        \\
WMLE(h=0.5) & 0.6276                        & 0.6260                        & {\color{red} 0.6463} & 0.6333                        \\
WMLE(h=1)   & 0.6447                        & 0.6236                        & 0.6382                        & 0.6355                        \\
WMLE(h=1.5) & 0.6341                        & 0.6065                        & 0.6171                        & 0.6192                        \\
Elo         & 0.6309                        & 0.6325                        & 0.6431                        & 0.6355                           \\ \hline
\end{tabular}
\end{table}

Somewhat surprisingly, our method achieved comparable results to Elo and outperformed Elo when $h=1$ in the average accuracy. This result suggests that season-length data is more informative for estimation and prediction. Note that Elo is a feature-rich method that uses more feature data to adjust the margin of victory than KRC. We also present the performance of the static rank centrality, which had the worst results in all seasons. Our method can balance long-term strength and local features, leading to better performance in forecasting for the entire season.

Table \ref{tab1} shows that the KRC method outperforms WMLE \citep{bong2020nonparametric} for most bandwidths. While the MLE is the most commonly used method, it may not be effective in practice, where the Bradley-Terry or the dynamic Bradley-Terry assumption violates due to external covariates. Factors such as injuries, referees, and others may impact game results, causing them to deviate from the dynamic Bradley-Terry model. The results presented in Table \ref{tab1} provide evidence that  KRC is more suitable to such data than the MLE and its dynamic extensions.

\begin {figure}[ht]
\centering
\includegraphics[width=0.8\textwidth]{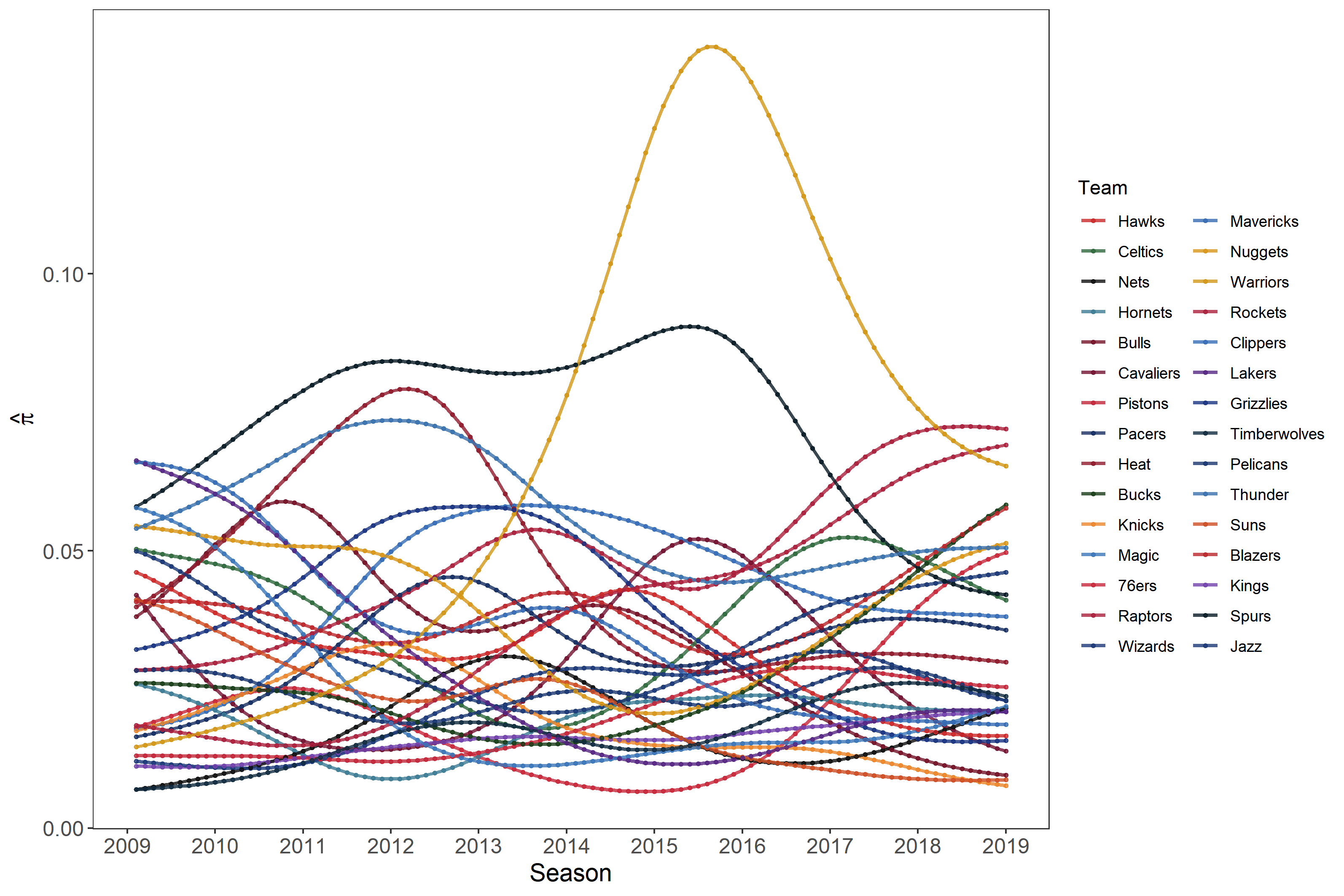}
\caption{Strength curves estimated for NBA teams. \label{fig_curve}}
\end{figure}

 We applied the KRC method to estimate the strength curve for each team. Figure \ref{fig_curve} illustrates the changes in team strength over ten seasons. As shown, the Warriors and Spurs are the two standout teams during these seasons. The Warriors won three championships in the last five seasons, while the Spurs maintained a high level of strength throughout the ten seasons. The Lakers and Magic were top teams in the early seasons, but their strengths decreased rapidly. These results are consistent with reality, indicating that the curves estimated by our method can validly reflect changes in team strengths.

In addition to estimating team strength, one may also wish to construct a confidence interval for the winning probability between pairs of interest. Table \ref{tab2} displays the estimated probability of one team winning over another at the end of the 2017-2018 season. The 95\% confidence interval uses the distribution theory of $\hat{\bm{\pi}}(t)$ and the delta method, with the bias term ignored in computation. Interestingly, the champion team of the season, the Warriors,  outperformed others with a probability higher than 0.5. However, the 95\% confidence interval with the Rockets team contains 0.5, suggesting that the Rockets were a strong opponent for the Warriors during the 2017-2018 season.

\begin{table}[!t]
\caption{Estimated paired winning probabilities, where each entry in the table presents an estimated probability of the team of the column beating the team of the row (with the 95\% confidence interval in the parentheses).   \label{tab2}}
\centering
\begin{tabular}{cccccc}
\hline
& Raptors      & Nets         & Warriors     & Knicks       & Pelicans     \\ \hline
\multirow{2}{*}{Celtics}   & 0.57         & 0.25         & 0.61         & 0.18         & 0.37         \\
& (0.51, 0.63) & (0.16, 0.33) & (0.55, 0.67) & (0.08, 0.28) & (0.30, 0.43) \\
\multirow{2}{*}{Mavericks} & 0.77         & 0.45         & 0.80         & 0.35         & 0.59         \\
& (0.68, 0.86) & (0.40, 0.51) & (0.70, 0.90) & (0.29, 0.42) & (0.51, 0.65) \\
\multirow{2}{*}{Rockets}   & 0.47         & 0.18         & 0.51         & 0.13         & 0.28         \\
& (0.42, 0.53) & (0.08, 0.29) & (0.46, 0.57) & (0.01, 0.25) & (0.20, 0.37) \\
\multirow{2}{*}{Lakers}    & 0.76         & 0.44         & 0.78         & 0.34         & 0.58         \\
& (0.67, 0.85) & (0.38, 0.49) & (0.69, 0.88) & (0.27, 0.41) & (0.52, 0.63) \\
\multirow{2}{*}{Pacers}    & 0.63         & 0.30         & 0.67         & 0.22         & 0.43         \\
& (0.57, 0.70) & (0.22, 0.37) & (0.60, 0.74) & (0.13, 0.27) & (0.37, 0.48) \\
\multirow{2}{*}{Cavaliers} & 0.75         & 0.43         & 0.78         & 0.33         & 0.57         \\
& (0.66, 0.84) & (0.37, 0.49) & (0.68, 0.88) & (0.26, 0.40) & (0.51, 0.63) \\ \hline
\end{tabular}
\end{table}

\section{Discussion}
\label{sec_diss}

This article introduces Kernel Rank Centrality (KRC), a spectral method devised to estimate item ratings in a dynamic setting based on the time-varying Bradley-Terry model. By integrating the Rank Centrality estimator with kernel smoothing techniques, KRC provides an accurate and efficient solution to the challenges posed by dynamic ranking. Our theoretical analysis of KRC demonstrates that its estimates converge to the dynamic Bradley-Terry model score with an $\ell_2$ error bound of $O_p(\sqrt{\frac{1}{nMh}}+h^2)$. This paper also establishes the asymptotic normality of KRC's distribution, a feat hitherto unachieved for spectral methods.

Our numerical investigation illuminates KRC's efficacy in sports statistics and validates the developed theory, showcasing its computational efficiency. Data analysis using NBA statistics indicates that our ranking method outperforms the esteemed Elo method in forecasting accuracy, underscoring KRC's effectiveness in dynamic rating. Thus, KRC can serve as a foundational tool for designing efficient algorithms for estimating and inferring sequential pairwise comparison data. Future research could explore time-dependent features and multi-dimensional comparison results to enhance KRC's performance.

\appendix

\section{Proofs of the results in Section \ref{sec_theo}}\label{appendix_A}
 In this section, we begin by demonstrating the uniqueness result, followed by an analysis of the $\ell_2$ convergence rate. Then, we provide the expansion result for our KRC estimator. Furthermore, we leverage the expansion to establish the $\ell_{\infty}$ result and the Central Limit Theorem (CLT) as direct applications.
 
	\subsection{Proof of Lemma \ref{lemma_unique}}
 \label{proof_a}
	The irreducibility of the chain $\hat{\bm{P}}(t)$ is equivalent to the strong connectivity of the directed graph with the adjacency matrix $\hat{\bm{P}}(t)$. That means if we divide the items into two nonempty subsets, there exists some team $i$ in the
	first set and some team $j$ in the second set satisfy $\hat{P}_{ij}(t)>0$. Note that for $(i,j)$, $\hat{P}_{ij}(t)>0$ if $j$ has beat $i$ one time in the $M_{ij}$ comparisons. Then a subset $\mathcal{Q}\subset [n]$, the probability that $\mathcal{Q}$ is not connected with $\mathcal{Q}^C$ is less than $(\frac{\kappa}{1+\kappa})^{M|\mathcal{Q}||\mathcal{Q}^C|}$. Let $\mathcal{W}$ denote the event that $\hat{\bm{P}}(t)$ is irreducible and $q=\frac{\log\frac{1}{2}}{\log\frac{\kappa}{1+\kappa}}+1$. Note that $\frac{\kappa}{1+\kappa}$ and $q$ are constants.
	Then, when $n$ and $M$ are sufficiently large, we have 
	\begin{align}
	\mathbb{P}(\mathcal{W})\geqslant& 1-\sum_{\mathcal{Q}\subset [n], |\mathcal{Q}|>0}(\frac{\kappa}{1+\kappa})^{M|\mathcal{Q}||\mathcal{Q}^C|}\nonumber\\
	\geqslant& 1-2\sum_{0<|\mathcal{Q}|\leqslant q}\binom{n}{|\mathcal{Q}|} (\frac{\kappa}{1+\kappa})^{M|\mathcal{Q}||\mathcal{Q}^C|}-\sum_{q<|\mathcal{Q}|<n-q}\binom{n}{|\mathcal{Q}|} (\frac{\kappa}{1+\kappa})^{M|\mathcal{Q}||\mathcal{Q}^C|}\nonumber\\
	\geqslant& 1-2n^q(\frac{\kappa}{1+\kappa})^{nM}-2^n(\frac{\kappa}{1+\kappa})^{Mq(n-q)} \nonumber\\
	\geqslant& 1-O((nM)^{-10}).
	\end{align}
	The last inequality holds with $2(\frac{\kappa}{1+\kappa})^q<1$ and $\frac{\kappa}{1+\kappa}<1$.
 
	\subsection{Proof of the $\ell_2$ rate in Theorem \ref{Theorem_1}}
       \label{proof_b}
 
	First we present the lemma of the perturbation bound for the leading left eigenvector of the probability transition matrix, a result developed in \cite{Negahban2017} and \cite{Chen2019_Spectral}.
	\begin{lemma}\label{lemma1}
		For two reversible Markov chains defined with transition matrix $\bm{P}$ and $\bm{P}^*$.  Suppose the stationary distributions are $\hat{\bm{\pi}}$ and $\bm{\pi}$ respectively. 
		Let $ \bm{E} =\bm{P}-\bm{P}^{*}$ and $\lambda_{\max }(\bm{P}^*)$ be the max absolute eigenvalue except $1$. If
		$1- \lambda_{\max }\left(\bm{P}^{*}\right)-\sqrt{\kappa}\|\bm{E}\|>0$,
		then 
		$$
		\left\|\hat{\bm{\pi}}-\bm{\pi}\right\| \leq \frac{\sqrt{k}\left\|\bm{\pi}^{\top} \bm{E}\right\|}{1- \lambda_{\max }\left(\bm{P}^{*}\right)-\sqrt{\kappa}\|\bm{E}\|}.
		$$
	\end{lemma}
	By virtue of Lemma \ref{lemma1}, the perturbation of the stationary distribution of the designed reversible Markov
	chain $\bm{P}^{*}(t)$ is dictated by two important quantities: (i) the condition number $1- \lambda_{\max }\left(\bm{P}^{*}(t)\right)$ and (ii) the noise size $\|\bm{E}(t)\|=\|\hat{\bm{P}}(t)-\bm{P}^{*}(t)\|$.
	The first part is controlled by the following lemma, which is a direct result induced from the proposition in static Rank Centrality \citep{Negahban2017}.
	\begin{lemma}\label{lemma2}
		If $\frac{\max_{i}\pi_i(t)}{\min_{i}\pi_i(t)}\leq \kappa$ and $\bm{P}^{*}(t)$ is the transition matrix defined in (\ref{f_P*}), we have
		$1-\lambda_{\max }\left(\bm{P}^{*}(t)\right) \geqslant \frac{1}{2 \kappa^{2}}.$
	\end{lemma}
	For the left part, we will show the noise size is well bounded by our kernel method with some mild conditions. For $i, j \in [n]$, let $\Delta_{ij}(t)$, $\bar{\Delta}_{ij}(t)$ and $\tilde{\Delta}_{ij}(t)$ denote $\frac{\sum_{t_k\in T_{ij}} y_{ij}(t_k)K_h(t,t_k)}{\sum_{t_k\in T_{ij}}K_h(t,t_k)}-y_{ij}^*(t)$, $\frac{\sum_{t_k\in T_{ij}} (y_{ij}(t_k)-y^*_{ij}(t_k))K_h(t,t_k)}{\sum_{t_k\in T_{ij}}K_h(t,t_k)}$ and $\frac{\sum_{t_k\in T_{ij}} (y_{ij}^*(t_k)-y^*_{ij}(t))K_h(t,t_k)}{\sum_{t_k\in T_{ij}}K_h(t,t_k)}$, respectively. 
	\begin{lemma} \label{lemma3}
		Under the conditions in Theorem \ref{Theorem_1}, there exists global constants $C_5$ and $C_6$, such that
		$$
		\mathbb{P}(\|\bm{E}(t)\|\leqslant C_5\sqrt{\frac{\log{(nMh)}}{nMh}}+C_6h^2)\rightarrow1.
		$$
	\end{lemma}

\paragraph*{Proof of Lemma \ref{lemma3}:}
		It is noted that the error matrix can be decomposed as
		$
		\|\bm{E}(t)\|\leqslant \|\bm{E}^D(t)\|+\|\bm{E}^U(t)\|+\|\bm{E}^L(t)\|,
		$
		where the superscript $D$, $U$ and $L$ denote the diagonal matrix, upper triangular matrix, and lower triangular matrix, respectively.
		Then we deal with these three terms separately.
		By the definition of the spectral norm, we know
		\begin{align}
		\|\bm{E}^D(t)\|
		= &\max_{i}\frac{1}{n}\left|\sum_{j\neq i}\sum_{t_m\in T_{ij}}\frac{K_h(t-t_{m})(y_{ij}(t_m)-y^*_{ij}(t))}{\sum_{t_m\in T_{ij}}K_h(t-t_{m})}\right|\nonumber\\
		\leqslant& \frac{1}{n}\max_{i}(|D_{1,i}|+|D_{2,i}|),
		\end{align}
		where,
		$$
		D_{1,i}=\sum_{j\neq i}\sum_{t_m\in T_{ij}}\frac{K_h(t-t_{m})(y_{ij}(t_m)-y^*_{ij}(t_m))}{\sum_{t_m\in T_{ij}}K_h(t-t_{m})}$$ 
  and 
  $$
		D_{2,i}=\sum_{j\neq i}\sum_{t_m\in T_{ij}}\frac{K_h(t-t_{m})(y_{ij}^*(t_m)-y^*_{ij}(t))}{\sum_{t_m\in T_{ij}}K_h(t-t_{m})}.
		$$
		
		$D_{1,i}$ is a sum of bounded independent random variables. Using Hoeffding's inequality, we have 
		\begin{align}
		\mathbb{P}\left(\left|D_{1,i}\right|\leqslant x\right)
		\geqslant 1- 2\exp{\left[-2x^2\sum_{j\neq i}\frac{(\sum_{t_m\in T_{ij}}K_h(t-t_{m}))^2}{\sum_{t_m\in T_{ij}}K_h^2(t-t_{m})}\right]}.  
		\nonumber 
		\end{align}
		When $M$ is large enough, we have 
		\begin{align}
		\sum_{j\neq i}\frac{\sum_{t_m\in T_{ij}}K_h^2(t-t_{m})}{(\sum_{t_m\in T_{ij}}K_h(t-t_{m}))^2}
		&=\sum_{j\neq i}\frac{\frac{1}{M_{ij}^2h^2}\sum_{t_m\in T_{ij}}K_h^2(t-t_{m})}{(\frac{1}{M_{ij}h}\sum_{t_m\in T_{ij}}K_h(t-t_{m}))^2}\nonumber\\&=\sum_{j\neq i}\frac{1}{M_{ij}h}\left(\frac{\int{K^2(v)\mathrm{d}v}}{\int{K(v)\mathrm{d}v}}+o(1)\right)=O\left(\frac{n}{Mh}\right).\nonumber
		\end{align}
		Let $x=\sqrt{\frac{2n\log{(nMh)}\int{K^2(v)\mathrm{d}v}}{Mh}}$ and we have that
		\begin{equation}
		\mathbb{P}\left(\left|D_{1,i}\right|\leqslant \sqrt{\frac{2n\log{(nMh)}\int{K^2(v)\mathrm{d}v}}{Mh}}\right)\geqslant 1-(nMh)^{-2}. \label{D1}
		\end{equation}
		
		Then we move on to $D_{2,i}$, which is the fluctuation part caused by the dynamic strength. As $Mh\rightarrow \infty$ and $h\rightarrow 0$,
			\begin{align}
	D_{2,i}&=\sum_{j\neq i}\frac{\int K(v)(y_{ij}^*(t+vh)-y_{ij}^*(t))\mathrm{d}v}{\int K(v)\mathrm{d}v+o(h^2)}\nonumber\\
	&=\sum_{j\neq i}\frac{\int K(v)(\dot{y}_{ij}^*(t)hv+\ddot{y}_{ij}^*(t)h^2v^2+O(h^3))\mathrm{d}v}{\int K(v)\mathrm{d}v+o(h^2)}\nonumber\\
	&=\sum_{j\neq i}\frac{h^2\ddot{y}_{ij}^*(t)\int v^2K(v)\mathrm{d}v+o(h^2)}{1+o(h^2)}.
	\end{align} 
		Let $\eta(t)=\max_{i,j}|\ddot{y}_{ij}^*(t)|$.
		We get, when $h$ is sufficiently small, 
		$$
		|D_{2,i}|\leqslant 2nh^2\eta(t)\int v^2K(v)\mathrm{d}v.$$ 
		Hence, together with (\ref{D1}), we complete the bound for $\|\bm{E}^D(t)\|$,
		\begin{align}
		\mathbb{P}\left(\|\bm{E}^D(t)\|\leqslant \sqrt{\frac{2\log{(nMh)}\int{K^2(v)\mathrm{d}v}}{nMh}}+2h^2\eta(t)\int v^2K(v)\mathrm{d}v\right)
		\geqslant 1-(nMh)^{-2}. \label{ED}
		\end{align}
		
		For the upper triangular part $\|\bm{E}^U(t)\|$, we apply the similar decomposition,
		$
		\|\bm{E}^U(t)\|\leqslant \|\bm{U}^{(1)}\|+\|\bm{U}^{(2)}\|,
		$
		where $\bm{U}^{(1)}$ and $\bm{U}^{(2)}$ are upper triangular matrices defined by
		$
		U^{(1)}_{ij}=\frac{1}{n}\bar{\Delta}_{ij}(t)$ and 
		$U^{(2)}_{ij}=\frac{1}{n}\tilde{\Delta}_{ij}(t).
		$
		With the result of $D_2$, we have
		$
		\|\bm{U}^{(2)}\|\leqslant  2h^2\eta(t)\int K(v)v^2{d}v. 
		$
		To deal with $\|\bm{U}^{(1)}\|$, we decompose it as a sum of independent random matrices,
		$
		\bm{U}^{(1)}=\sum_{i<j}\sum_{t_m\in T_{ij}}\bm{Z}^{(ijm)}.
		$
		$\bm{Z}^{(ijm)}$ is an $n\times n$ matrix whose elements are zeros except ${Z}^{(ijm)}_{ij}=\frac{1}{n}\frac{K_h(t-t_{m})(y_{ij}^*(t_m)-y^*_{ij}(t))}{\sum_{t_m\in T_{ij}}K_h(t-t_{m})}$.
		Then we can bound these independent matrices by, 
		\begin{equation}
		\|{Z}^{(ijm)}\|\leqslant \frac{1}{n}\frac{K_h(t-t_{m})}{\sum_{t_m\in T_{ij}}K_h(t-t_{m})}\leqslant \frac{2\max_v K_(v)}{nMh}.
		\end{equation}
		The second inequality holds with the fact that
		$\frac{\sum_{t_m\in T_{ij}}K_h(t-t_{m})}{Mh}\geqslant\frac{1}{2}$ when $M$ is sufficiently large. Note that the elements in $\bm{U}^{(1)}$ are independent.
		We also have
		\begin{align}
		\|\mathbb{E} [\bm{U}^{(1)^\top}\bm{U}^{(1)}]\|= \max_{i}{\mathbb{E}[ \bm{U}^{(1)}_{i\cdot}\bm{U}_{i\cdot}^{(1)^\top}}]=\max_{i}\sum_{j>i}\sum_{t_m\in T_{ij}}\mathbb{E}[\bm{Z}_{ij}^{(ijm)2}]
		\leqslant \frac{1}{3nMh}\frac{\int{K^2(v)\mathrm{d}v}}{\int{K(v)\mathrm{d}v}}.\nonumber
		\end{align}
		Then applying the matrix Bernstein’s inequality \citep{tropp2012user}, it holds that for any $x>0$,
		\begin{align}
		\mathbb{P}&(\|\bm{U}^{(1)}\| \leqslant x)
		\geqslant 1-2n\exp{\left[\frac{-x^2}{\frac{1}{6nMh}\frac{\int{K^2(v)\mathrm{d}v}}{\int{K(v)\mathrm{d}v}}+\frac{\max_v K_(v)}{3nMh}x}\right]}.\nonumber
		\end{align}
		Let $x=\sqrt{\frac{2\log{(nMh)}\int{K^2(v)\mathrm{d}v}}{nMh}}$ and combining $\|\bm{U}^{(2)}\|$. Then
		\begin{align}
		\mathbb{P}\left(\|\bm{E}^U(t)\|\leqslant \sqrt{\frac{2\log{(nMh)}\int{K^2(v)\mathrm{d}v}}{nMh}}+2h^2\eta(t)\int K(v)v^2{d}v\right)\geqslant 1-n^{-5}(Mh)^{-6}. 
\end{align}
		The similar arguments lead to the same upper bound for $\|\bm{E}^L(t)\|$,
		which we omit for brevity. Finally, let $\eta=\sup_{t \in [0,1]} \eta(t)$ and we complete the proof by combining the constant multipliers in these bounds.

\paragraph*{Proof of $\ell_2$ rate in Theorem \ref{Theorem_1}:}
		By the bounds of $\|\bm{E}(t)\|$ and $1-\lambda_{max}(\bm{P}^*(t))$, we have that $1-\lambda_{max}(\bm{P}^*(t))-\sqrt{\kappa}\|\bm{E}(t)\|\geqslant\frac{1}{4\kappa^2}$ holds with probability at least $1-(nMh)^{-2}$ , when $\sqrt{\frac{\log{(nMh)}}{nMh}}+h^2$ is sufficiently small.
		Then we only need to demonstrate the bound $\|\bm{\pi}(t)^\top\bm{E}(t)\|$
		to complete the proof of $\ell_2$ rate in Theorem \ref{Theorem_1}.
		For simplicity of discussion, we decompose $\bm{E}(t)$ into $\bar{\bm{E}}(t)=\hat{\bm{P}}(t)-\mathbb{E}(\hat{\bm{P}}(t))$ and $\tilde{\bm{E}}(t)=\mathbb{E}(\hat{\bm{P}}(t))-\bm{P}^*(t)$. With the discussion in the proof of Lemma \ref{lemma3}, we know $\|\tilde{\bm{E}}(t)\|\lesssim h^2$ with a global constant. As a result, 
		\begin{equation}
		\|\bm{\pi}(t)^\top\tilde{\bm{E}}(t)\|\lesssim h^2\|\bm{\pi}(t)\|. \label{f_piEh}
		\end{equation}	
		
		So we focus on $\|\bm{\pi}(t)^\top\bar{\bm{E}}(t)\|$.
		Note that $\bar{\bm{E}}(t)$ is constructed by $\{\bar{\Delta}_{ab}(t)\}_{1\leqslant a<b\leqslant n}$. Then $\|\bm{\pi}(t)\bar{\bm{E}}(t)\|^2$ is in a quadratic form of $\bar{\bm{\Delta}}(t)$, the $\frac{n(n-1)}{2}$ dimensional column vector from $\{\bar{\Delta}_{ab}(t)\}_{1\leqslant a<b\leqslant n}$.
		Specifically,
		$$
		\|\bm{\pi}(t)^\top\bar{\bm{E}}(t)\|^2=\bm{\pi}(t)^\top\bar{\bm{E}}(t)\bar{\bm{E}}(t)^\top\bm{\pi}(t)=\bar{\bm{\Delta}}(t)^\top\bm{G}\bm{G}^\top\bar{\bm{\Delta}}(t), 
		$$
		where $\bm{G}$ is an $\frac{n(n-1)}{2}\times n$ matrix. For convenience, we use the subscript $ab$ denote the row or column corresponding to $\bar{\Delta}_{ab}(t)$. The matrix $\bm{G}$ is defined by
		\begin{equation}\label{f_G}
		G_{ab,i}=\left\{
		\begin{array}{lll}
		\displaystyle -\frac{1}{n}(\pi_a+\pi_b) & \text{if $i= a$,}   \\
		\displaystyle \frac{1}{n}(\pi_a+\pi_b) & \text{if  $i = b$,}\\
		\displaystyle 0& \text{otherwise.}
		\end{array}
		\right.
		\end{equation}
		Then $\bm{G}^\top\bm{G}$ is a weighted Laplacian matrix with $(\bm{G}^\top\bm{G})_{kk}=\frac{1}{n^2}\sum_{l\neq k} (\pi_k+\pi_l)^2$ and $(\bm{G}^\top\bm{G})_{kl}=-\frac{1}{n^2}(\pi_k+\pi_l)^2$. Then by the spectrum theory of Laplacian matrices \citep{Bojan1991}, we can obtain
		\begin{align}
		n\min_{i,j}\frac{1}{n^2}(\pi_i+\pi_j)^2\lesssim\|\bm{G}^\top\bm{G}\|\lesssim n\max_{i,j}\frac{1}{n^2}(\pi_i+\pi_j)^2.\nonumber
		\end{align}
		Hence, $\|\bm{G}\bm{G}^\top\|=\|\bm{G}^\top\bm{G}\|=O(\frac{1}{n^3})$. And the Frobenius norm of $\bm{G}\bm{G}^\top$ can be bounded by
		$$
		\|\bm{G}\bm{G}^\top\|_F\leqslant \sqrt{\mathrm{Rank}(\bm{G}\bm{G}^\top)}\|\bm{G}\bm{G}^\top\|
		\lesssim \frac{1}{\sqrt{n^5}}.
		$$
		Note that $\bar{\Delta}_{ab}(t)$ is subgaussian.
		Then by Hanson-Wright inequality \citep{HW2013}, we can derive, for some constant $c>0$,
		\begin{align}
		&\mathbb{P}(|\bar{\bm{\Delta}}(t)^\top\bm{G}\bm{G}^\top\bar{\bm{\Delta}}(t)-\mathbb{E}[\bar{\bm{\Delta}}(t)^\top\bm{G}\bm{G}^\top\bar{\bm{\Delta}}(t)]|\geqslant x)\nonumber\\>&1-\exp(-c\min(\frac{(Mh)^2x^2}{\|\bm{G}\bm{G}^\top\|_F^2},\frac{Mhx}{\|\bm{G}\bm{G}^\top\|})).
		\end{align}
		
		This together with $\mathbb{E}[\bar{\bm{\Delta}}(t)^\top\bm{G}\bm{G}^\top\bar{\bm{\Delta}}(t)]\lesssim \frac{1}{nMh}\|\bm{\pi}(t)\|^2$, we have with probability at least $1-(nMh)^{-2}$,
		\begin{align}
		\|\bm{\pi}(t)^\top\bar{\bm{E}}(t)\|\lesssim \sqrt{\frac{\|\bm{\pi}(t)\|^2}{nMh}+\frac{\log(nMh)}{\sqrt{n^5}Mh}}\lesssim \|\bm{\pi}(t)\|\sqrt{\frac{1}{nMh}}\label{f_piE}.
		\end{align}
		Note that the inequalities (\ref{f_piEh}) and (\ref{f_piE}) hold with global constants. So combine the bounds, and we get to the $\ell_2$ result.	
		The $\ell_{\infty}$ convergence rate in Theorem \ref{Theorem_1} is the result obtained from the bounds of $B_1$ and $B_2$ in the proof of Theorem \ref{theorem_expansion}.

	\subsection{Expansion results for the spectral method}
 \label{proof_c}
	
	We first present a non-asymptotic expansion for the KRC method, which is a direct result from the Maclaurin expansion for an irreducible Markov chain \citep{cao_1998}.
	
	\begin{lemma}[Maclaurin expansion]
		Let $\bm{E}(t)=\hat{\bm{P}}(t)-\bm{P}^*(t)$ and $\bm{A}^{\#}(t)$ denote the group inverse of $\bm{A}(t)=\bm{I}-\bm{P}^*(t)$. 
		If $\|\bm{E}(t)\bm{A}^{\#}(t)\|<1$, then
		\begin{align}\label{f_Maclaurins}
			\hat{\bm{\pi}}(t)^\top-\bm{\pi}(t)^\top=
			\bm{\pi}(t)^\top\bm{E}(t)\bm{A}^{\#}(t)+\hat{\bm{\pi}}(t)^\top\bm{E}(t)\bm{A}^{\#}(t)\bm{E}(t)\bm{A}^{\#}(t).
		\end{align}
	\end{lemma}

	To analyze the asymptotics of $\hat{\bm\pi}(t)$ as $n \rightarrow \infty$, we need 
	to know the structure of $\bm{A}^{\#}(t)$. One way is to approximate $\bm{A}^{\#}(t)$ by $\tilde{\bm{A}}(t)$ in a
	simple form such that
	$$
	\hat{\pi}_i(t)-\pi_i(t)\approx \bm{\pi}(t)^\top\bm{E}(t)\bm{A}^{\#}_{.i}(t)\approx \bm{\pi}(t)^\top\bm{E}(t)\tilde{\bm{A}}_{.i}(t).
	$$
	
	To achieve this, we observe that the group inverse matrix can be approximated by a diagonal matrix with the max column $\ell_{2}$ error bounded by $O(\frac{1}{\sqrt{n}})$. 
	\begin{theorem}[Group inverse approximation]\label{theorem_gm}
		Let $\tilde{\bm{A}}(t)$ be a diagonal matrix with the diagonal elements defined by
		$$\tilde{A}_{ii}(t)=\frac{1}{A_{ii}(t)}=\frac{1}{\sum_{j\neq i}\frac{1}{n}y^*_{ij}(t)}.$$
		Then when $\kappa=O(1)$, we have, 
		$$\max_{i\in [n]}\|\tilde{\bm{A}}_{.i}(t)-\bm{A}^{\#}_{.i}(t)\|=O\left(\frac{1}{\sqrt{n}}\right).$$
	\end{theorem}
	
	The approximation is in a sharp and simple form such that the asymptotic property is easy to analyze. With this approximation, we simplify the expansion in \eqref{f_Maclaurins} in a more concise and tractable form. 
	
	Next, we express the entrywise expansion as a sum of these pairwise error terms.
	
	\begin{theorem}[Entrywise expansion]\label{theorem_expansion}
		\label{expansion}
		Under the assumptions of Theorem \ref{Theorem_1},  for any $i\in[n]$ and $t \in (0,1)$, if $n \rightarrow \infty$, $Mh\rightarrow \infty$ and $nMh^7\rightarrow 0$ , we have,  with probability tending to $1$,
		\begin{align}
			\hat{\pi}_i(t)-\pi_i(t)=&\frac{1}{\sum_{j\neq i} y_{ij}^*(t)}\sum_{j\neq i}(\pi_i(t)+\pi_j(t))\bar\Delta_{ij}(t)+\beta_{i}(t)h^2+\epsilon_{i}(t)\nonumber
			,
		\end{align}
		where $\displaystyle \sup_{t\in(0,1)} \max_{i}|\beta_i(t)|=O(\frac{1}{n})$ and $\displaystyle \sup_{t\in(0,1)} \max_{i}|\epsilon_i(t)|=o_p\left(\sqrt{\frac{1}{n^3Mh}}\right)$.
		
	\end{theorem}

	The proof of this expansion involves substituting $\tilde{\bm{A}}_{.i}(t)$ into (\ref{f_Maclaurins}) and bounding the remaining terms. A similar expansion result is also derived for the static setting in recent work by \cite{gao2021uncertainty}. However, their proof relies on a result of $\ell_{\infty}$ convergence rate \citep{Chen2019_Spectral} and the leave-one-out trick. For our method, the $\ell_{\infty}$ convergence rate is a direct consequence of this expansion, making our approach more straightforward. Using this expansion, we can easily derive the $\ell_{\infty}$ convergence rate and the finite-dimensional asymptotic normality for the $\hat{\bm{\pi}}(t)$ in the previous subsection.

\paragraph*{Proof of Theorem \ref{theorem_gm}:}
		By the property of the Markov chain and the group inverse, we have
		$
		(\bm{A}(t)+\bm{e}\bm{\pi}(t)^\top)\bm{A}^{\#}(t)=\bm{I}-\bm{e}\bm{\pi}(t)^\top.
		$
		Then for $i\in [n]$,
		\begin{align}
		(\bm{A}(t)+\bm{e}\bm{\pi}(t)^\top)(\bm{A}_{.i}^{\#}(t)-\tilde{\bm{A}}_{.i}(t))=\bm{I}_{.i}-\bm{e}\pi_i(t)-(\bm{A}(t)+\bm{e}\bm{\pi}(t)^\top)\tilde{\bm{A}}_{.i}(t).
		\end{align}
		In the right side, by direct calculations, it is shown that
		\begin{align}
		\bm{A}_{i.}(t)\tilde{\bm{A}}_{.i}(t)&=1, \nonumber\\
		\bm{A}_{j.}(t)\tilde{\bm{A}}_{.i}(t)&=-\frac{1}{\sum_{j\neq i}y^*_{ij}(t)}\frac{\pi_i(t)}{\pi_i(t)+\pi_j(t)}, j\neq i,\\
		\bm{e}\bm{\pi}(t)^\top\tilde{\bm{A}}_{.i}(t)&=\bm{e}\pi_i(t)\frac{1}{\sum_{j\neq i}\frac{1}{n}y^*_{ij}(t)}.\nonumber
		\end{align}
		Let $\bm{R}=\bm{I}_{.i}-\bm{e}\pi_i(t)-(\bm{A}(t)-\bm{e}\bm{\pi}(t)^\top)\tilde{\bm{A}}_{.i}(t)$.
		Combine these parts and we have 
		$
		R_i=(-1-\frac{1}{\sum_{j\neq i}\frac{1}{n}y^*_{ij}(t)})\pi_i(t)$ and for $j\neq i$, 
		$
		R_j=(-1+\frac{1}{\sum_{j\neq i}y^*_{ij}(t)}\frac{1}{\pi_i(t)+\pi_j(t)}
		-\frac{1}{\sum_{j\neq i}\frac{1}{n}y^*_{ij}(t)})\pi_i(t)
		.$
		
		Note that $\frac{1}{\sum_{j\neq i}y^*_{ij}(t)}\frac{1}{\pi_i(t)+\pi_j(t)}$ and $\frac{1}{\sum_{j\neq i}\frac{1}{n}y^*_{ij}(t)}$ are $O(1)$ when $\kappa$ is $O(1)$. Hence $\|\bm{R}\|=O(\|\bm{e}\pi_i(t)\|)=O(\frac{1}{\sqrt{n}})$.
		For the left side, 
		\begin{align}
		\|(\bm{A}(t)+\bm{e}\bm{\pi}(t)^\top)(\bm{A}_{.i}^{\#}(t)-\tilde{\bm{A}}_{.i}(t))\|
		\geqslant\sigma_{min}(\bm{A}(t)+\bm{e}\bm{\pi}(t)^\top)\|\bm{A}_{.i}^{\#}(t)-\tilde{\bm{A}}_{.i}(t)\|,
		\end{align}
		where $\sigma_{min}(\bm{A}(t)+\bm{e}\bm{\pi}(t)^\top)$ is the minimum singular value of $\bm{A}(t)+\bm{e}\bm{\pi}(t)^\top$. It is noted that $\sigma_{min}(\bm{A}(t)+\bm{e}\bm{\pi}(t)^\top) \asymp 1-\lambda_{max}(\bm{P}^{*}(t))$ when $\kappa$ is $O(1)$. With the result of Lemma \ref{lemma2}, we obtain that
		$
		\|\bm{R}\| \gtrsim \frac{1}{2\kappa^2}\|\bm{A}_{.i}^{\#}(t)-\tilde{\bm{A}}_{.i}(t)\|.
	    $
		Together with $\|\bm{R}\|=O(\frac{1}{\sqrt{n}})$, we get that $\|\bm{A}_{.i}^{\#}(t)-\tilde{\bm{A}}_{.i}(t)\|$ is $O(\frac{1}{\sqrt{n}})$. 

\paragraph*{Proof of Theorem \ref{theorem_expansion}:}
	When $\|\bm{E}(t)\bm{A}^{\#}(t)\|<1$, for any $i\in [n]$,
		$$
		\hat{\pi}_i(t)-\pi_i(t)=\bm{\pi}(t)^\top\bm{E}(t)\bm{A}^{\#}_{.i}(t)
		+\hat{\bm{\pi}}(t)^\top\bm{E}(t)\bm{A}^{\#}(t)\bm{E}(t)\bm{A}^{\#}_{.i}(t).
		$$
		
		By the discussion in the proof of Theorem \ref{Theorem_1}, the condition holds with probability tending to one.
		The first term is the main term in the expansion. For simplicity, we omit the notation of time $t$ when there is no ambiguity.
		Plugging the approximation in the first term of the expansion, we derive the entrywise expansion for $\hat{\pi}_i(t)$. For $i\in [n]$,
		\begin{align}
		&\hat{\pi}_i(t)-\pi_i(t)\nonumber\\
		=&\frac{1}{n}\sum_{k < l}(\pi_k+\pi_l)(A^{\#}_{li}-A^{\#}_{ki})\Delta_{kl}(t)+\hat{\bm{\pi}}^\top\bm{E}\bm{A}^{\#}\bm{E}\bm{A}^{\#}_{.i}\nonumber\\
		=&\frac{1}{n}\sum_{k < l}(\pi_k+\pi_l)(\tilde{A}_{li}-\tilde{A}_{ki})\Delta_{kl}(t)\nonumber
		\\
		&+\frac{1}{n}\sum_{k < l}(\pi_k+\pi_l)(A^{\#}_{li}-\tilde{A}_{li}+\tilde{A}_{ki}-A^{\#}_{ki})\Delta_{kl}(t)\nonumber\\
		&+\hat{\bm{\pi}}^\top\bm{E}\bm{A}^{\#}\bm{E}\bm{A}^{\#}_{.i}.
		\end{align}

		We denote the three parts in right side of the expansion as $B_1$, $B_2$ and $B_3$. $B_1$ is tractable due to the simple form of $\tilde{\bm{A}}(t)$,
		\begin{align}
		B_1=\frac{1}{\sum_{j\neq i} y_{ij}^*}\sum_{j\neq i}(\pi_i(t)+\pi_j(t))\Delta_{ij}(t).
		\end{align}
		Using Hoeffding’s inequality and the similar technique in bounding $\|E^D(t)\|$, we see that $B_1$ is $O_p(\frac{1}{n}(\sqrt{\frac{\log(nMh)}{nMh}}+h^2))$. Then we need to prove that the remaining parts are $o_p(\frac{1}{n}(\sqrt{\frac{1}{nMh}}+h^2))$.
		Note that each element of $B_2$ is bounded and we have
		\begin{align} \label{f_B2b}
		&\sum_{k < l} [(\pi_k(t)+\pi_l(t))(A^{\#}_{li}-\tilde{A}_{li}+\tilde{A}_{ki}-A^{\#}_{ki})]^2 \nonumber\\
		\leqslant& 4\max_{j\in[n]}{\pi_j(t)^2}\sum_{k < l }2[(A^{\#}_{li}-\tilde{A}_{li})^2+(\tilde{A}_{ki}-A^{\#}_{ki})^2]\nonumber\\
		\leqslant& 8n\max_{j\in[n]}{\pi_j(t)^2}\|\tilde{A}_{.i}-A^{\#}_{.i}\|^2 \lesssim \frac{1}{n^2}.
		\end{align} 	
		Then use the decomposition again,
				\begin{align}
		B_2=
		&\frac{1}{n}\sum_{k < l}\bigg[(\pi_k+\pi_l)(A^{\#}_{li}-\tilde{A}_{li}+\tilde{A}_{ki}-A^{\#}_{ki})\bar\Delta_{kl}(t)\bigg]
		\nonumber\\
		&+\frac{1}{n}\sum_{k < l}\bigg[(\pi_k+\pi_l)(A^{\#}_{li}-\tilde{A}_{li}+\tilde{A}_{ki}-A^{\#}_{ki})\tilde\Delta_{kl}(t)\bigg]\nonumber\\
		:=&B_{21}+B_{22}.
		\end{align}
		For $B_{21}$, we have 
		\begin{align}
		&\sum_{k < l} [(\pi_k(t)+\pi_l(t))(A^{\#}_{li}-\tilde{A}_{li}+\tilde{A}_{ki}-A^{\#}_{ki})]^2\frac{\sum\limits_{t_m}K_h^2(t-t_m)}{(\sum\limits_{t_m}K_h(t-t_m))^2}\nonumber\\
		\lesssim&\sum_{k < l} [(\pi_k(t)+\pi_l(t))(A^{\#}_{li}-\tilde{A}_{li}+\tilde{A}_{ki}-A^{\#}_{ki})]^2\frac{1}{Mh}\lesssim \frac{1}{n^2Mh}.
		\end{align}
		The last inequality is obtained with the bound in (\ref{f_B2b}).
		By Hoeffding’s inequality, we can derive that $B_{21}$ is $O_p(\frac{1}{n^2}\sqrt{\frac{\log(Mn)}{Mh}})$.
		For the bias term $B_{22}$,
		\begin{align}
		|B_{22}|
		&\lesssim 
		\frac{1}{n} \sqrt{\sum_{k < l } [(\pi_k(t)+\pi_l(t))(A^{\#}_{li}-\tilde{A}_{li}+\tilde{A}_{ki}-A^{\#}_{ki})]^2h^4}
		\lesssim \frac{h^2}{n}.
		\end{align}
		The first inequality follows from Cauchy-Schwartz's inequality and the fact that the bias for each element is $O(h^2)$. Combining the bound of $B_{22}$ and the bound of the bias term in $B_1$, we can obtain that $|\beta_i(t)|=O(\frac{1}{n})$.

		$B_3$ is the second order remainder. To handle this term, we decouple the dependence of $\hat{\bm{\pi}}(t)$ and $E(t)$ by replacing $\hat{\bm{\pi}}(t)$ with $\bm{\pi}(t)$, 
		\begin{align}
		B_3=\bm{\pi}^{\top}\bm{E}\bm{A}^{\#}\bm{E}\bm{A}^{\#}_{.i}+(\hat{\bm{\pi}}-\bm{\pi})^\top\bm{E}\bm{A}^{\#}\bm{E}\bm{A}^{\#}_{.i}:=B_{31}+B_{32}.\nonumber
		\end{align}
		By Theorem 1, $\|\hat{\bm{\pi}}-\bm{\pi}\|$ is $O_p(\sqrt{\frac{1}{n}}(\sqrt{\frac{1}{Mnh}}+h^2))$ and $\|\bm{E}\|$ is $O_p(\sqrt{\frac{\log(nMh)}{nMh}}+h^2)$. Then by the norm property,
		\begin{align}
		B_{32}\leqslant \|\hat{\bm{\pi}}-\bm{\pi}\|\|\bm{E}\|\|\bm{A}^{\#}\|\|\bm{E}\|\|\bm{A}^{\#}\|.
		\end{align}
Note that $\|\bm{A}^{\#}\|\asymp\frac{1}{1-\lambda_{max}(\bm{P}^{*}(t))}\asymp 1$ when $\kappa=O(1)$. Then, $B_{32}=o_p(\frac{1}{n}(\sqrt{\frac{1}{nMh}}+h^2))$.
		
		For $B_{31}$, we substitute the decomposition of $\bm{E}$,
		\begin{align}
		\bm{\pi}^{\top}\bm{E}\bm{A}^{\#}\bm{E}\bm{A}^{\#}_{.i}=&\bm{\pi}^{\top}(\bar{\bm{E}}+\tilde{\bm{E}})\bm{A}^{\#}(\bar{\bm{E}}+\tilde{\bm{E}})\bm{A}^{\#}_{.i}\nonumber\\
		=&\bm{\pi}^{\top}\bar{\bm{E}}\bm{A}^{\#}\bar{\bm{E}}\bm{A}^{\#}_{.i}+
		\bm{\pi}^{\top}\bar{\bm{E}}\bm{A}^{\#}\tilde{\bm{E}}\bm{A}^{\#}_{.i}\nonumber\\
		&+
		\bm{\pi}^{\top}\tilde{\bm{E}}\bm{A}^{\#}\bar{\bm{E}}\bm{A}^{\#}_{.i}+
		\bm{\pi}^{\top}\tilde{\bm{E}}\bm{A}^{\#}\tilde{\bm{E}}\bm{A}^{\#}_{.i}
		\end{align}
Denote the four parts as $F_1$, $F_2$, $F_3$ and $F_4$, respectively, where
$F_4$ is tractable for analysis. We can analyze the $\ell_{\infty}$ norm for the multiplicative product step by step.
		Let $\tilde{\bm{E}}^D$ be the diagonal matrix of $\tilde{\bm{E}}$.
		With the facts $\|\tilde{\bm{E}}^D\|_{\infty}=O(h^2)$ and $\|\tilde{\bm{E}}^D\|_{\infty}=O(\frac{h^2}{n})$, we have
		$
		\|\bm{\pi}^{\top}\tilde{\bm{E}}\|_{\infty}\leqslant \|\bm{\pi}^{\top}\tilde{\bm{E}}^D\|_{\infty}+\|\bm{\pi}^{\top}(\tilde{\bm{E}}-\tilde{\bm{E}}^D)\|_{\infty}\lesssim \frac{h^2}{n}.
		$
		For each $j\in [n]$,
		\begin{align}
		\bm{\pi}^{\top}\tilde{\bm{E}}\bm{A}^{\#}_{.j}&\leqslant |(\bm{\pi}^{\top}\tilde{\bm{E}})_{j}A^{\#}_{jj}|+|\sum_{k\neq j}(\bm{\pi}^{\top}\tilde{\bm{E}})_{k}A^{\#}_{kj}|\nonumber\\
		&\leqslant \|\bm{\pi}^{\top}\tilde{\bm{E}}\|_{\infty}|A^{\#}_{jj}|+\sqrt{n}\|\bm{A}^{\#}_{(-j)j}\|\|\bm{\pi}^{\top}\tilde{\bm{E}}\|_{\infty} \nonumber\\
		&\lesssim\frac{h^2}{n}.
		\end{align}
		Decompose $\tilde{\bm{E}}$ again and we have
		\begin{align}
		\|\bm{\pi}^{\top}\tilde{\bm{E}}\bm{A}^{\#}\tilde{\bm{E}}\|_{\infty}
		\leqslant& \|\bm{\pi}^{\top}\tilde{\bm{E}}\bm{A}^{\#}\tilde{\bm{E}}^D\|_{\infty}+
		\|\bm{\pi}^{\top}\tilde{\bm{E}}\bm{A}^{\#}(\tilde{\bm{E}}-\tilde{\bm{E}}^D)\|_{\infty}\nonumber\\
		\leqslant &
		\|\bm{\pi}^{\top}\tilde{\bm{E}}\bm{A}^{\#}\|_{\infty}\|\tilde{\bm{E}}^D\|_{\infty}+n\|\bm{\pi}^{\top}\tilde{\bm{E}}\bm{A}^{\#}\|_{\infty}\|\tilde{\bm{E}}-\tilde{\bm{E}}^D\|_{\infty}\nonumber\\
		\lesssim &\frac{h^4}{n}.\label{f_step3}
		\end{align}
		We decompose $\bm{A}^{\#}_{.i}$ again in the last step,
		\begin{align}
		F_4
		\leqslant \|\bm{\pi}^{\top}\tilde{\bm{E}}\bm{A}^{\#}\tilde{\bm{E}}\|_{\infty}|\bm{A}^{\#}_{ii}|+\|\bm{\pi}^{\top}\tilde{\bm{E}}\bm{A}^{\#}\tilde{\bm{E}}\|\|\bm{A}^{\#}_{(-i)i}\|
		\lesssim \frac{h^4}{n}.\label{f_step4}
		\end{align}		
		When $nMh^7\rightarrow 0$, $F_4$ is $o(\sqrt{\frac{1}{n^3Mh}})$.
		These steps will be helpful when we bound $F_2$ and $F_3$. For $F_2$, from the previous result, we know $\|\bm{\pi}^{\top}\bar{\bm{E}}\bm{A}^{\#}\|_{\infty}=O_p(\sqrt{\frac{\log(nMh)}{n^3Mh}})$. So use the same decomposition method in (\ref{f_step3}) and (\ref{f_step4}), and we derive that $F_2=O_p(h^2\sqrt{\frac{\log(nMh)}{n^3Mh}})$.		
		Note that $\|\bm{\pi}^{\top}\tilde{\bm{E}}\bm{A}^{\#}\|_{\infty}=O(\frac{h^2}{n})$. We can derive the bound for $(\bm{\pi}^{\top}\tilde{\bm{E}}\bm{A}^{\#}\bar{\bm{E}}\bm{A}^{\#}_{.i})$ in the same manner as analyzing $\bm{\pi}^{\top}\bar{\bm{E}}\bm{A}^{\#}_{.i}$. Hence, we also have $F_3=O_p(h^2\sqrt{\frac{\log(nMh)}{n^3Mh}})$. When $nMh^7\rightarrow 0$ and $h\rightarrow 0$, $h^2\sqrt{\log(nMh)}=o(1)$. So we have $F_2$ and $F_3$ are both $o_p(\sqrt{\frac{1}{n^3Mh}})$.
		Then for $F_{1}$, it can be written as the quadratic form of $\bar{\bm{\Delta}}$,
		\begin{align}
		F_1=\bm{\pi}^{\top}\bar{\bm{E}}\bm{A}^{\#}\bar{\bm{E}}\bm{A}^{\#}_{.i}=\bar{\bm{\Delta}}^\top\bm{G}\bm{A}^{\#}\bm{H}\bar{\bm{\Delta}},
		\end{align}
		where $\bm{G}$ is an $\frac{n(n-1)}{2}\times n$ matrix and $\bm{H}$ is an $n\times\frac{n(n-1)}{2}$ matrix. 		
		Next we will present the bounds for the spectral norm and Frobenius norm of $\bm{G}\bm{A}^{\#}\bm{H}$. It is noted that  $\bm{G}$ is the same matrix defined in (\ref{f_G}) and $\|\bm{G}\|=O(\frac{1}{\sqrt{n^3}})$.
		The discussion for $H$ is similar with that of $\bm{G}$. $\bm{H}\bm{H}^\top$ has the same structure with $\|\bm{G}^\top\bm{G}\|$. By replacing $(\pi_k+\pi_l)$ with $(A^{\#}_{ki}+A^{\#}_{li})$, we have
	    $
		\|\bm{H}\bm{H}^\top\|\lesssim n\max_{k,l}\frac{1}{n^2}(A^{\#}_{ki}+A^{\#}_{li})^2.$
		So the order for $\||\bm{H}\|$ is $\sqrt{\frac{1}{n}}$. Combine the two bounds and we can control the spectral norm of $\bm{G}\bm{A}^{\#}\bm{H}$ by
		\begin{align}
		Q=\|\bm{G}\bm{A}^{\#}\bm{H}\|\leqslant\|\bm{G}\|\|\bm{A}^{\#}\|\|\bm{H}\|\lesssim\frac{1}{n^2}.
		\end{align}
		For simplicity, we denote $\bm{G}\bm{A}^{\#}\bm{H}$ by $\bm{Q}$.
		To handle the Frobenius norm, we derive the explicit form of $\bm{Q}$. In fact, $\bm{Q}$ is the second derivative matrix of $\hat{\bm{\pi}}(t)$ with respect to $\bm{y}^*(t)$. For the pairs $ab$ $(a<b)$ and $cd$ $(c<d)$, the second order derivative is in the following form, 
		\begin{align}
		\bm{Q}_{ab,cd}
		=&\bm{\pi}(t)(\frac{\partial\bm{P}^*(t)}{\partial y^*_{ab}(t)}\bm{A}^{\#}\frac{\partial\bm{P}^*(t)}{\partial y^*_{cd}(t)}+\frac{\partial\bm{P}^*(t)}{\partial y^*_{cd}(t)}\bm{A}^{\#}\frac{\partial\bm{P}^*(t)}{\partial y^*_{ab}(t)})\bm{A}^{\#}_{.i}\nonumber\\
		=&\frac{1}{n^2}(\pi_a(t)+\pi_b(t))(A^{\#}_{ac}-A^{\#}_{bc}+A^{\#}_{ad}-A^{\#}_{bd})(A^{\#}_{ci}-A^{\#}_{di})\nonumber\\
		&+\frac{1}{n^2}(\pi_c(t)+\pi_d(t))(A^{\#}_{ca}-A^{\#}_{da}+A^{\#}_{cb}-A^{\#}_{db})(A^{\#}_{ai}-A^{\#}_{bi}).
		\end{align}
		So we can bound the Frobenius norm by 
		\begin{align}
		\|\bm{Q}\|_F\lesssim \frac{1}{n^2}\|\bm{\pi}(t)\|_{\infty}\sqrt{n}\|\bm{A}^{\#}\|_F\|\bm{A}^{\#}{.i}\|=O(\frac{1}{n^2}).
		\end{align}
		Then we discuss the expectation of $B_{31}$.
		\begin{align}
		\mathbb{E}&[\bar{\bm{\Delta}}^\top\bm{Q}\bar{\bm{\Delta}}]=\sum_{ab,cd}Q_{ab,cd}\mathbb{E}[\bar{\Delta}_{ab}\bar{\Delta}_{cd}]
		=\sum_{ab}Q_{ab,ab}\mathbb{E}[\bar{\Delta}_{ab}\bar{\Delta}_{ab}]
		\nonumber\\
		&\lesssim \sum_{ab}Q_{ab,ab}\frac{1}{Mh}
	\lesssim \frac{1}{n^3}n\|\bm{A}^{\#}\|_{\infty}\|\bm{A}^{\#}_{.i}\|_1 \frac{1}{Mh}%
		\lesssim \frac{1}{n^2Mh}.\nonumber
		\end{align}
		The last inequality is from the facts that
		$
		\|\bm{A}^{\#}_{.i}\|_1\leqslant |\tilde{\bm{A}}_{.i}|_1+ |\bm{A}^{\#}_{.i}-\tilde{\bm{A}}_{.i}|_1
		\leqslant |\tilde{\bm{A}}_{.i}|_1+ \sqrt{n}|\bm{A}^{\#}_{.i}-\tilde{\bm{A}}_{.i}|_2\lesssim 1,
		$
		and 
		$
		\|\bm{A}^{\#}\|_{\infty}\leqslant \max_{i\in[n]}\|\bm{A}^{\#}_{.i}\|_1\lesssim 1.
		$
		Then by the Hanson-Wright inequality \citep{HW2013}, we can derive that $F_1$ is $O_p(\frac{1}{n^2Mh})$.
		Finally, combine the bounds for $F_1$, $F_2$, $F_3$, $F_4$ and we prove that
		$B_{31}$ is $o_p(\frac{1}{n}(\sqrt{\frac{1}{Mnh}}+h^2))$.

	\subsection{Proof of Theorem \ref{theorem_clt}}
		\label{proof_d}
		By the expansion, we have for any $i \in [n]$, with probability tending to 1,
		\begin{align}
		&\hat{\pi}_i(t)-\pi_i(t)-\beta_i(t)h^2\nonumber\\
		=&\frac{1}{\sum_{j\neq i} y_{ij}^*(t)}\sum_{j\neq i}(\pi_i(t)+\pi_j(t))\bar{\Delta}_{ij}(t)+o\left(\sqrt{\frac{1}{n^3Mh}}\right).
		\nonumber
		\end{align}
		
		Let $\mathcal{A}$ denote the event that the expansion holds true. We first show the conditional asymptotic normality under $\mathcal{A}$.
		Let $f_i$ denote the main term in the expansion.  It suffices to prove that for any fixed $\bm{\gamma} \in \mathbb{R}^s$ and $\|\bm{\gamma}\|=1$,
		$
		\sum_{i=1}^s  \gamma_i\alpha_i(t)f_i\buildrel d\over\longrightarrow N(0,1).
		$
		Note that
		$$
		\mathbb{E}[f_i^2]=\frac{1}{(\sum_{j\neq i} y_{ij}^*(t))^2}\sum_{j\neq i}(\pi_i(t)+\pi_j(t))^2\mathbb{E}[\bar{\Delta}_{ij}^{2}(t)],
		$$
		and 
		$$
		\mathbb{E}[f_if_j]=\frac{1}{(\sum_{s\neq i} y_{is}^*(t))(\sum_{s\neq j} y_{js}^*(t))}(\pi_i(t)+\pi_j(t))^2\mathbb{E}[\bar{\Delta}_{ij}^{2}(t)].
		$$
		Then 
		\begin{align}
		\mathbb{E}&\left[(\sum_{i=1}^s  \gamma_i\alpha_i(t) f_i)^2\right] =\sum_{i=1}^{s}\gamma_i^2\alpha_i^2(t)\mathbb{E}[f_i^2]\nonumber\\
		&\qquad+2\sum_{1\leqslant i<j\leqslant s}\frac{\gamma_i\gamma_j(\pi_i(t)+\pi_j(t))^2\mathbb{E}[\bar{\Delta}_{ij}^{2}(t)]}{\sqrt{\sum_{q\neq i}(\pi_i(t)+\pi_q(t))^2\mathbb{E}[\bar{\Delta}_{iq}^{2}(t)]}\sqrt{\sum_{q\neq j}(\pi_j(t)+\pi_q(t))^2\mathbb{E}[\bar{\Delta}_{jq}^{2}(t)]}}.
		\end{align}
		It is noted that each term in the second term is $O(\frac{1}{n})$. Hence,
		$$
		\mathbb{E}\left[(\sum_{i=1}^s  \gamma_i\alpha_i(t) f_i)^2\right]\rightarrow \sum_{i=1}^{s}\gamma_i^2\alpha_i^2(t)\mathbb{E}[f_i^2]+0=1.
		$$
		Moreover,  the Lindeberg-Feller condition holds trivially since $\{\bar{\Delta}_{ij}(t)\}$ are bounded.
		So by Lindeberg-Feller Central Limit Theorem, $\hat{\pi}_i(t)-\pi_i(t)-\bm{\Gamma}_{i.}(t)\bm{b}(t)h^2|\mathcal{A} \buildrel d\over\longrightarrow N(0,1)$.
		The proof is complete with the fact that for every $x \in \mathbb{R}$,
		\begin{align}
		&\mathbb{P}(\hat{\pi}_i(t)-\pi_i(t)-\beta_i(t)h^2\leqslant x)\nonumber\\
		=& \mathbb{P}(\hat{\pi}_i(t)-\pi_i(t)-\beta_i(t)h^2\leqslant x|\mathcal{A})\mathbb{P}(\mathcal{A})+\mathbb{P}(\hat{\pi}_i(t)-\pi_i(t)-\beta_i(t)h^2\leqslant x|\mathcal{A}^C)\mathbb{P}(\mathcal{A}^C)\nonumber\\
		\rightarrow&\mathbb{P}(\hat{\pi}_i(t)-\pi_i(t)-\bm{\Gamma}_{i.}(t)\bm{b}(t)h^2\leqslant x|\mathcal{A})+0. \nonumber
		\end{align}

\bibliographystyle{imsart-nameyear}
\bibliography{ref}

\end{document}